\documentclass[review,3p,times, 11pt]{elsarticle}   

\usepackage{enumitem}
\usepackage{lineno}
\usepackage{multirow}
\usepackage{rotating}
\usepackage{epstopdf}
\usepackage{graphicx}
\usepackage{subfigure}
\usepackage{times}

\usepackage{makeidx}
\usepackage{multirow}
\usepackage{multicol}
\usepackage{graphicx}
\usepackage{epstopdf}
\usepackage{ulem}

\usepackage{array}
\usepackage{amsmath,amssymb,amsfonts}
\usepackage{algorithmic}
\usepackage{algorithm}
\usepackage{graphicx}
\usepackage{textcomp}

\usepackage{epsfig}
\usepackage{makecell}
\usepackage{longtable,booktabs}

\usepackage{longtable}

\usepackage[colorlinks,
linkcolor=black,       
anchorcolor=black,  
citecolor=black,        
]{hyperref}

\biboptions{numbers,sort&compress}

\journal{Elsevier}

\begin{document}
\begin{frontmatter}
	
	
		
	\title{KCoreMotif: An Efficient Graph Clustering Algorithm for Large Networks by Exploiting \textit{k}-core Decomposition and Motifs}


\author[label1]{Gang Mei\corref{cor1}}
\ead{gang.mei@cugb.edu.cn}
\author[label1]{Jingzhi Tu}
\author[label1]{Lei Xiao}
\author[label2]{Francesco Piccialli\corref{cor1}}
\ead{francesco.piccialli@unina.it}
\cortext[cor1]{Corresponding author}
\address[label1]{School of Engineering and Technology, China University of Geosciences (Beijing), 100083, Beijing, China}
\address[label2]{Department of Mathematics and Applications R. Caccioppoli, University of Naples Federico II, Naples, Italy}

\begin{abstract}
Clustering analysis has been widely used in trust evaluation on various complex networks such as wireless sensors networks and online social networks. Spectral clustering is one of the most commonly used algorithms for graph-structured data (networks). However, the conventional spectral clustering is inherently difficult to work with large-scale networks due to the fact that it needs computationally expensive matrix manipulations. To deal with large networks, in this paper, we propose an efficient graph clustering algorithm, KCoreMotif, specifically for large networks by exploiting \textit{k}-core decomposition and motifs. The essential idea behind the proposed clustering algorithm is to perform the efficient motif-based spectral clustering algorithm on \textit{k}-core subgraphs, rather than on the entire graph. More specifically, (1) we first conduct the \textit{k}-core decomposition of the large input network; (2) we then perform the motif-based spectral clustering for the top \textit{k}-core subgraphs; (3) we group the remaining vertices in the rest (\textit{k}-1)-core subgraphs into previously found clusters; and finally obtain the desired clusters of the large input network. To evaluate the performance of the proposed graph clustering algorithm KCoreMotif, we use both the conventional and the motif-based spectral clustering algorithms as the baselines and compare our algorithm with them for 18 groups of real-world datasets. Comparative results demonstrate that the proposed graph clustering algorithm is accurate yet efficient for large networks, which also means that it can be further used to evaluate the intra-cluster and inter-cluster trusts on large networks.  
\end{abstract}

\end{frontmatter}


\newpage

\section{Introduction}
\label{sec1}

\begin{figure}[!ht]
	\centering
	\includegraphics[width=0.9\textwidth]{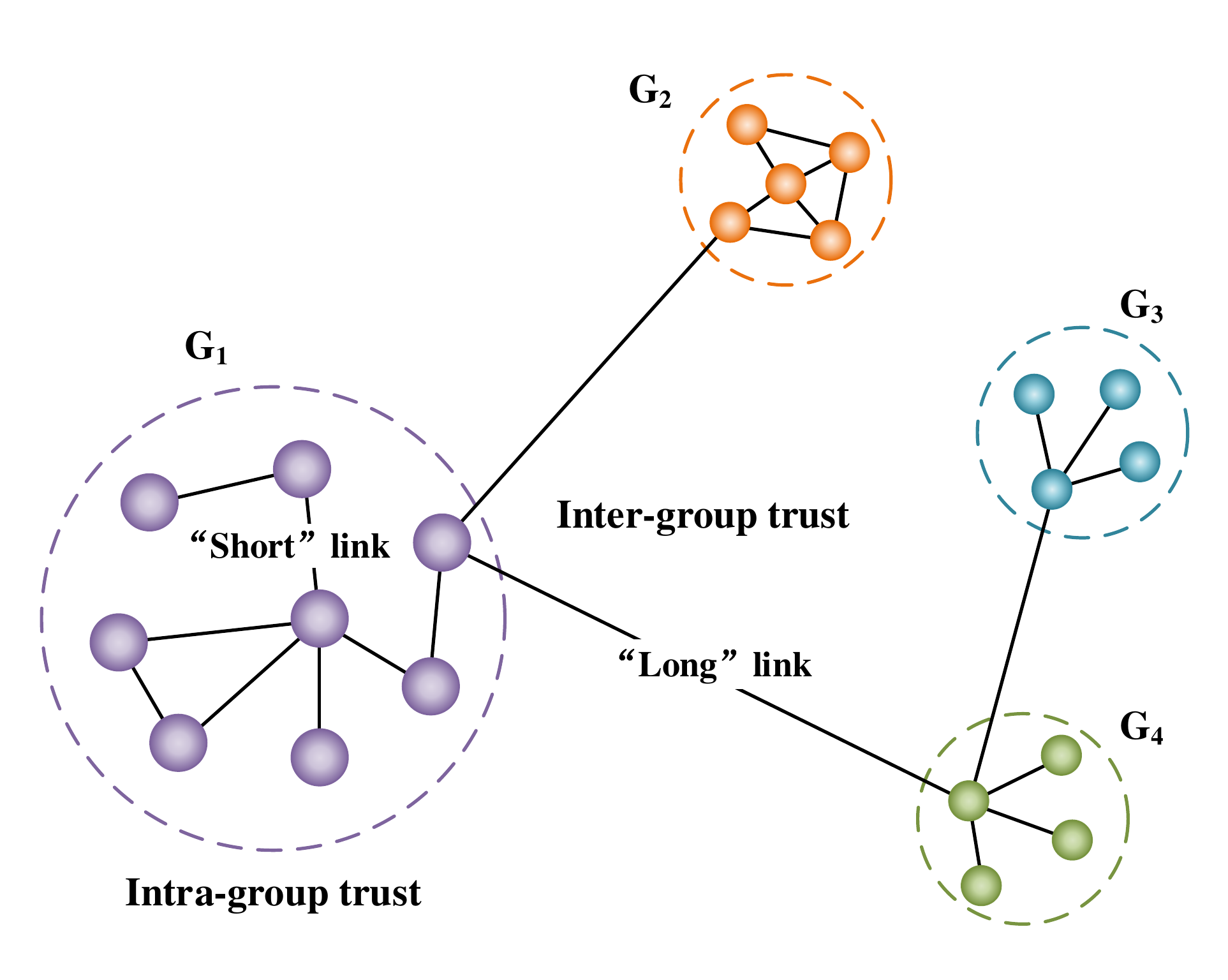}
	\caption{A simple illustration of trust evaluation within intra-clusters/groups and between inter-clusters/groups} 
	\label{Figure1}
\end{figure}

Graph Clustering analysis has been widely used in trust evaluation on various complex networks such as online social networks \cite{r01,r02,r03} and mobile Ad Hoc networks \cite{r04,r05}. Graph clustering intends to group vertices according to their characteristics such that there is high intra-cluster similarity and low inter-cluster similarity. After the graph clustering, the trust within intra-clusters/groups and between inter-clusters/groups could be analyzed and evaluated based on the clustered results; see a simple illustration in Figure \ref{Figure1}.

There are many methods for conducting graph clustering \cite{r06,r07,r08}. Spectral clustering \cite{r09} is one of the most commonly used algorithms for graph-structured data (also called \textit{networks}), which is capable of obtaining satisfying clustering results in most cases for graphs. The spectral clustering algorithm constructs a matrix, solves an associated eigenvalue problem, and extracts splitting information from the calculated eigenvectors.

However, the spectral clustering is inherently difficult to deal with large-scale networks. This is because that, in the spectral clustering, it needs to conduct computationally expensive matrix manipulations, e.g., (1) the construction of a large adjacency matrix and Laplacian matrix and (2) the calculation of eigenvectors of the large Laplacian matrix. Even if the large adjacency matrix and Laplacian matrix are stored in compressed and sparse formats such as the COOrdinate (COO) or Compressed Sparse Column (CSC) formats, the manipulations of matrices are quite computationally expensive. 

To address the problem inherently arising in the spectral clustering algorithm, much research work has been carried out to efficiently find clusters of large networks. Those efforts can be roughly classified into two categories: (1) designing parallel spectral clustering algorithms; and (2) reducing the computational cost of the matrix assembly and manipulation in the spectral clustering algorithm. 

The first strategy for conducting spectral clustering on large networks is to design parallel spectral algorithms on various platforms. For example, Huo, et al \cite{r10} designed an efficient parallel spectral clustering algorithm on multi-core processors in the Julia language \cite{r11}. Jin and Jaja \cite{r12} developed a high-performance implementation of spectral clustering on CPU-GPU Platforms. Chen, et al. \cite{r13} proposed the parallel spectral clustering in distributed systems. Jin, et al. \cite{r14} designed an efficient parallel spectral clustering algorithm for large datasets under the cloud computing environment. 

The second strategy for conducting spectral clustering on large networks is to reduce the computational cost of the matrix assembly and manipulation. For example, when clustering large-scale datasets, a popular method is to use the Nystr\"{o}m extension method \cite{r15} for approximating similarity matrix and feature decomposition to avoid directly calculating the similarity matrix and feature decomposition of the entire dataset. 

Nystr\"{o}m extension method was originally used to solve large-scale matrix eigendecomposition problems. In recent years, it has been successfully applied to large-scale machine learning algorithms, such as spectral clustering. Nystr\"{o}m extension method approximates the properties of the whole data set through a small number of sampling data. Therefore, the quality of the sampling data strongly affects the quality of the Nystr\"{o}m method.

The selection of sample points is critical in the Nystr\"{o}m extension method. One of the simplest sampling methods is random sampling. However, the results of random sampling are usually unstable. The performance of the Nystr\"{o}m extension method can be improved by selecting an effective sampling strategy, that is, selecting more meaningful sample points and using a small number of sample points to obtain satisfactory performance. At present, a lot of work has been conducted based on Nystr\"{o}m extension-based sampling strategies, such as the sampling algorithm based on \textit{k}-means center points \cite{r16}, sampling based on Schur complement \cite{r17}, incremental sampling algorithm based on variance \cite{r18}, incremental sampling method based on prediction error analysis \cite{r18}, selective sampling strategy \cite{r19}, adaptive partial sampling \cite{r20}, etc. These sampling algorithms include two kinds of algorithms: one is a one-time sampling method, and the other is the incremental sampling method. 

Another effective idea for reducing the computational cost is only to perform the spectral clustering on local subgraphs, rather than on the global graph. For example, in the research work introduced by Peng, et al \cite{r21} and Giatsidis, et al \cite{r22}, the global graph is first decomposed into several \textit{k}-core subgraphs, and then spectral clustering is conducted on the specified \textit{k}-subgraph. Then, the remaining vertices in the rest (\textit{k}-1) subgraphs are further grouped into the existing clusters that were previously determined by the spectral clustering algorithm. In such methods, spectral clustering is only conducted for subgraphs and thus reduces the computational cost. 

Recently, Leskovec, et al. \cite{r23} proposed an efficient motif-based spectral clustering method for large networks. The essential idea behind the motif-based spectral clustering is to cluster the nodes of a graph-based on motifs instead of edges, i.e., to find the higher-order organization of complex networks at the level of small network subgraphs (motifs) \cite{r23}. 

In this paper, we propose an efficient graph clustering algorithm, KCoreMotif, for large networks by exploiting \textit{k}-core decomposition and motifs. The essential idea behind the proposed clustering algorithm is to perform the motif-based spectral clustering algorithm on \textit{k}-core subgraphs. More specifically, (1) we first conduct the \textit{k}-core decomposition of the large input network; (2) we then perform the motif-based spectral clustering for the top \textit{k}-core Subgraphs; (3) we group the remaining vertices in the rest \textit{k}-core subgraphs into previously found clusters; and finally obtain the clusters of the large input network. To evaluate the performance of the proposed graph clustering algorithms, we use both the conventional and motif-based spectral clustering algorithms as the baseline and compare our algorithm with it for 18 groups of real-world datasets. 

Our contributions in this work can be summarized as follows.

(1) We design an efficient graph clustering algorithm (KCoreMotif) by exploiting \textit{k}-core decomposition and motifs for large networks. 

(2) We suggest the selection of the value of \textit{k} in the \textit{k}-core decomposition.

(3) We analyze the advantages and shortcomings of the proposed graph clustering algorithm and point out potential work for improving it. 

The rest of this paper is organized as follows. Section 2 will present the background introduction to the conventional spectral clustering algorithm. Section 3 will describe the details of the proposed graph clustering algorithm. Section 4 will list the experimental setup and results for evaluating the performance of the proposed graph clustering algorithm. Section 5 will discuss the proposed algorithm. Finally, Section 6 draws several conclusions for this work.

\section{Background: Conventional Spectral Clustering Algorithm}
\label{sec2}

In this section, we will briefly introduce the basic concepts and essential ideas behind the conventional spectral clustering algorithm. The spectral clustering is a graph theory-based unsupervised learning algorithm with high adaptability and superior clustering performance \cite{r24}. It uses the Laplacian matrix \cite{r25} to reduce the dimensionality of the datasets and then clusters the results according to eigenvectors. The spectral clustering algorithm can be applied to find clustering for both scattered data and graph-structured data.

\begin{figure}[!ht]
	\centering
	\includegraphics[width=\textwidth]{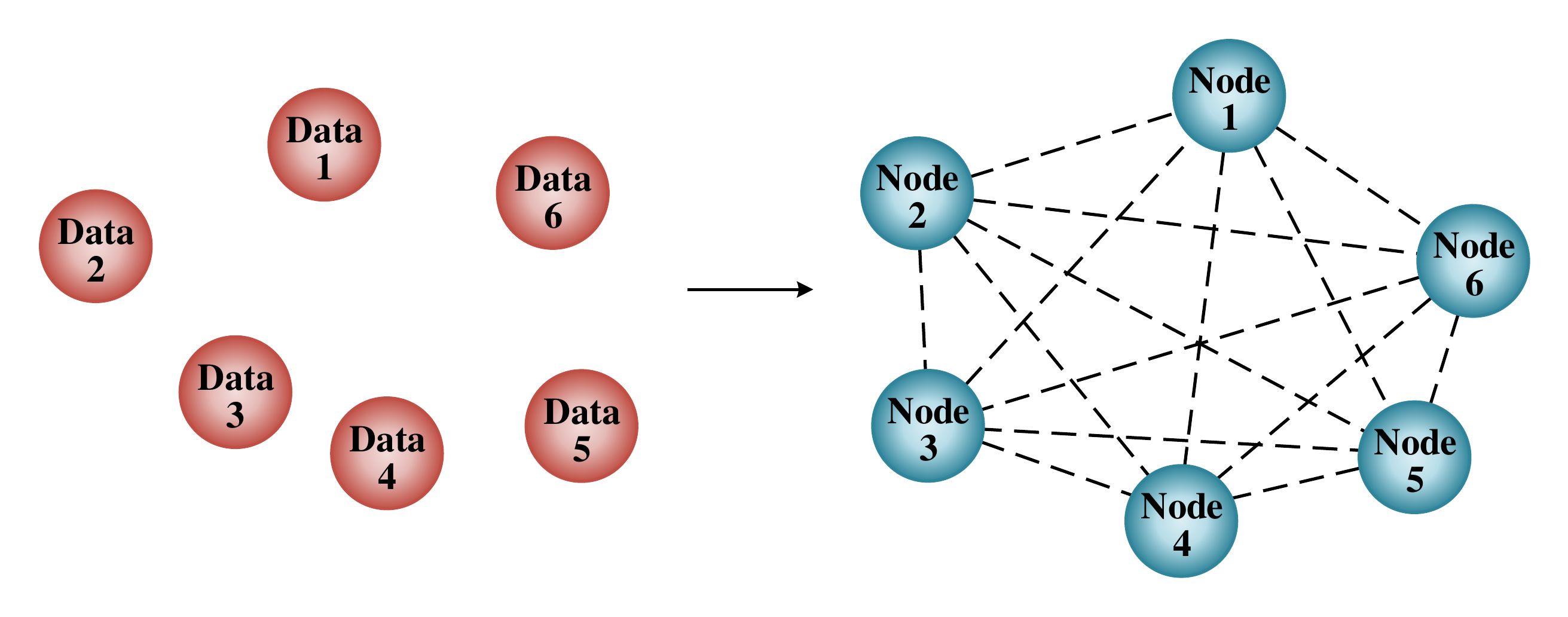}
	\caption{Forming the undirected graph \textit{G} from the scattered data} 
	\label{Figure2}
\end{figure}

When dealing with the scattered data, the data needs to first form a weighted undirected graph \textit{G}:

\begin{equation}
\label{eq1}
{G=(V,E), E=\left\{(i,j),S_{i,j}>0\right\}\subseteq V \times V}
\end{equation}

\noindent where \textit{V} is node set and \textit{E} is the edge set of the graph \textit{G}; see Figure \ref{Figure2}. 

The matter of clustering shifts to partitioning the graph into two or more optimal subgraphs so that the subgraphs are as similar as possible internally and the subgraphs are separated from each other by a large distance:

\begin{equation}
\label{eq2}
{cut(A,B)=\sum_{u\in A,v\in B}^{}W_{u,v}}
\end{equation}

\noindent where graph \textit{G} is split into two disjoint subsets of \textit{A} and \textit{B}.

There are many algorithms of undirected graph partitioning, such as Normalized cut (\textit{Ncut}) \cite{r26}, Average cut \cite{r27}, Minimum cut \cite{r28}, Ratio cut \cite{r29}, and Min-max cut \cite{r30}, of which \textit{Ncut} is the most prevalent one:

\begin{equation}
\label{eq3}
{Ncut(A,B)=\frac{cut(A,B)}{assoc(A,V)}+\frac{cut(A,B)}{assoc(B,V)}}
\end{equation}

\noindent where $ \textit{assoc}(A,V) = \sum_{u\in A,t\in V}^{}W_{u,v}$ and $ \textit{assoc}(B,V) = \sum_{u\in B,t\in V}^{}W_{u,v}$ are the connection from their nodes to all nodes in the graph.

In general, the process of the conventional spectral clustering algorithm can be summarized as the following four steps.

\textbf{Step 1. Constructing the adjacency matrix}

\begin{figure}[!ht]
	\centering
	\includegraphics[width=\textwidth]{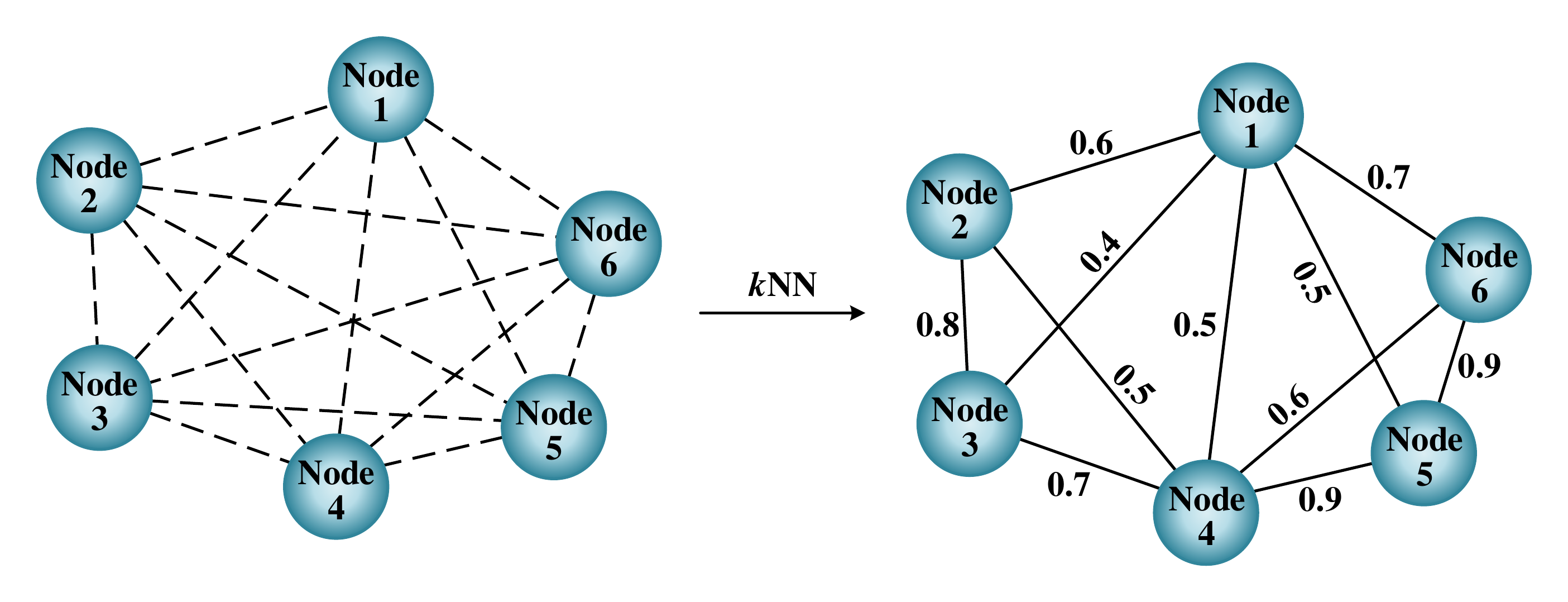}
	\caption{Constructing the adjacency matrix using the \textit{k}NN search} 
	\label{Figure3}
\end{figure}

$ \epsilon$-neighborhood, \textit{k}-nearest neighbors (\textit{k}NN), and fully connected are three different methods to construct the adjacency matrix \textit{W}. Here we take the \textit{k}NN search algorithm as an example (see Figure \ref{Figure3}):

\begin{equation}
\label{eq4}
{W_{i,j}=W_{j,i}=\begin{cases}0 & if \ x_{i}\notin kNN(x_{j})\ and\ x_{j}\notin kNN(x_{i})    \\e^{\frac{-\parallel x_{i}-x_{j}\parallel }{2\sigma^{2}}^{2}} & if\ x_{i}\in kNN(x_{j}) \ or\ x_{j}\in kNN(x_{i}) \end{cases}}
\end{equation}

\begin{equation}
\label{eq5}
{W_{i,j}=W_{j,i}=\begin{cases}0 & if \ x_{i}\notin kNN(x_{j})\ or \ x_{j}\notin kNN(x_{i})    \\e^{\frac{-\parallel x_{i}-x_{j}\parallel }{2\sigma^{2}}^{2}} & if\ x_{i}\in kNN(x_{j}) \ and\ x_{j}\in kNN(x_{i}) \end{cases}}
\end{equation}

These two algorithms ensure that the adjacency matrix is symmetric.

\textbf{Step 2. Constructing the Laplacian matrix}

The simplest form of the Laplacian matrix is as follows:

\begin{equation}
\label{eq6}
{L=D-W}
\end{equation}

\noindent where \textit{D} is the degree matrix, and \textit{W} is the weighted adjacency matrix.

The simplest Laplacian matrix can be further normalized to be symmetric. The construction of the normalized Laplacian matrix $ L_{sym}$ requires the introduction of degree matrix \textit{D}:

\begin{equation}
\label{eq7}
{D_{i,j}=\begin{cases}0 & if\ i\neq j\\\sum_{j}^{}W_{i,j} & if\ i=j \end{cases}}
\end{equation}

\noindent where $ W_{i,j}$ is the element in the adjacency matrix, and $ \sum_j^{} W_{i,j}$ is the sum of the weights of the edges that connect the node to other nodes.

Then we can obtain  $ L_{sym}$ :

\begin{equation}
\label{eq8}
{L_{sym}=D^{-\frac{1}{2}}\times L\times D^{-\frac{1}{2}}}
\end{equation}

\noindent where \textit{L} = \textit{D} - \textit{W}.

\textbf{Step 3. Computing the eigenvectors matrix}

Based on the desired number of clusters \textit{k}, we then find the first \textit{k} eigenvectors \{$ u_{1}$, $ u_{2}$,..., $ u_{k}$\} of $ L_{sym}$. Constructing the matrix \textit{U} of eigenvectors as columns and then normalizing matrix \textit{U}:

\begin{equation}
\label{eq9}
{U_{i,j}=\frac{U_{i,j}}{\sqrt{\sum_{k}^{}U_{i,k}^2}}}
\end{equation}

\textbf{Step 4. Conducting the \textit{k}-means clustering}

\begin{figure}[!ht]
	\centering
	\includegraphics[width=\textwidth]{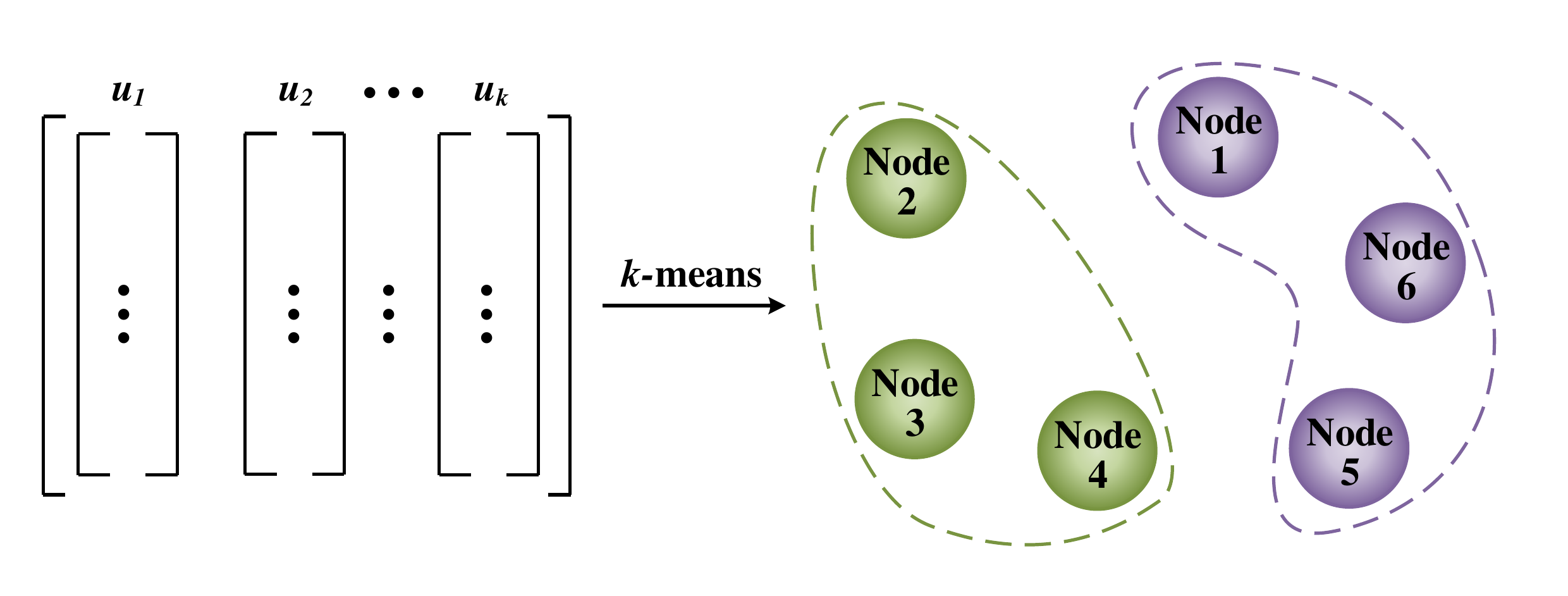}
	\caption{Clustering the eigenvectors matrix \textit{U}} 
	\label{Figure4}
\end{figure}

Each row in the matrix \textit{U} corresponds to each node in the graph and is clustered in columns. In this step, \textit{k}-means is replaceable by other clustering algorithms, such as \textit{k}-means++ \cite{r31}, which is an enhancement algorithm on the \textit{k}-means; see Figure \ref{Figure4}.

\section{Methods}
\label{sec3}
\subsection{Overview of the Proposed Graph Clustering Algorithm}

In this paper, we propose an efficient graph clustering algorithm for large networks by exploiting \textit{k}-core decomposition and motifs. We term the proposed graph clustering algorithm as KCoreMotif. The essential idea behind the proposed clustering algorithm is to perform the motif-based spectral clustering algorithm on \textit{k}-core subgraphs. The essential idea introduced in this paper is the utilization of the one introduced in \cite{r21,r22} and the other one introduced in \cite{r23}.

The objective of proposing the graph clustering algorithm is to work with large networks. The spectral clustering algorithm is one of the popular and effective methods for graph clustering, but it is quite computationally expensive for large networks. We attempt to utilize the advantages of the spectral clustering (i.e., it is effective to find clusters of graph-structured data), while providing an efficient solution to working with large networks by exploiting the \textit{k}-core decomposition and network motifs.

There are two essential ideas behind our algorithm to work with large networks:

(1)	We perform the spectral clustering locally, rather than globally. That is, the expensive spectral clustering is conducted on \textit{k}-core subgraphs, rather than on the entire graph. This idea may help reduce the computational cost of the spectral clustering.

(2)	We employ the motif-based spectral clustering algorithm, rather than the conventional spectral clustering algorithm. The motif-based clustering algorithm is much more efficient than the conventional one due to the replacement of the edge adjacency matrix with the motif adjacency matrix.

\begin{figure}[!ht]
	\centering
	\includegraphics[width=\textwidth]{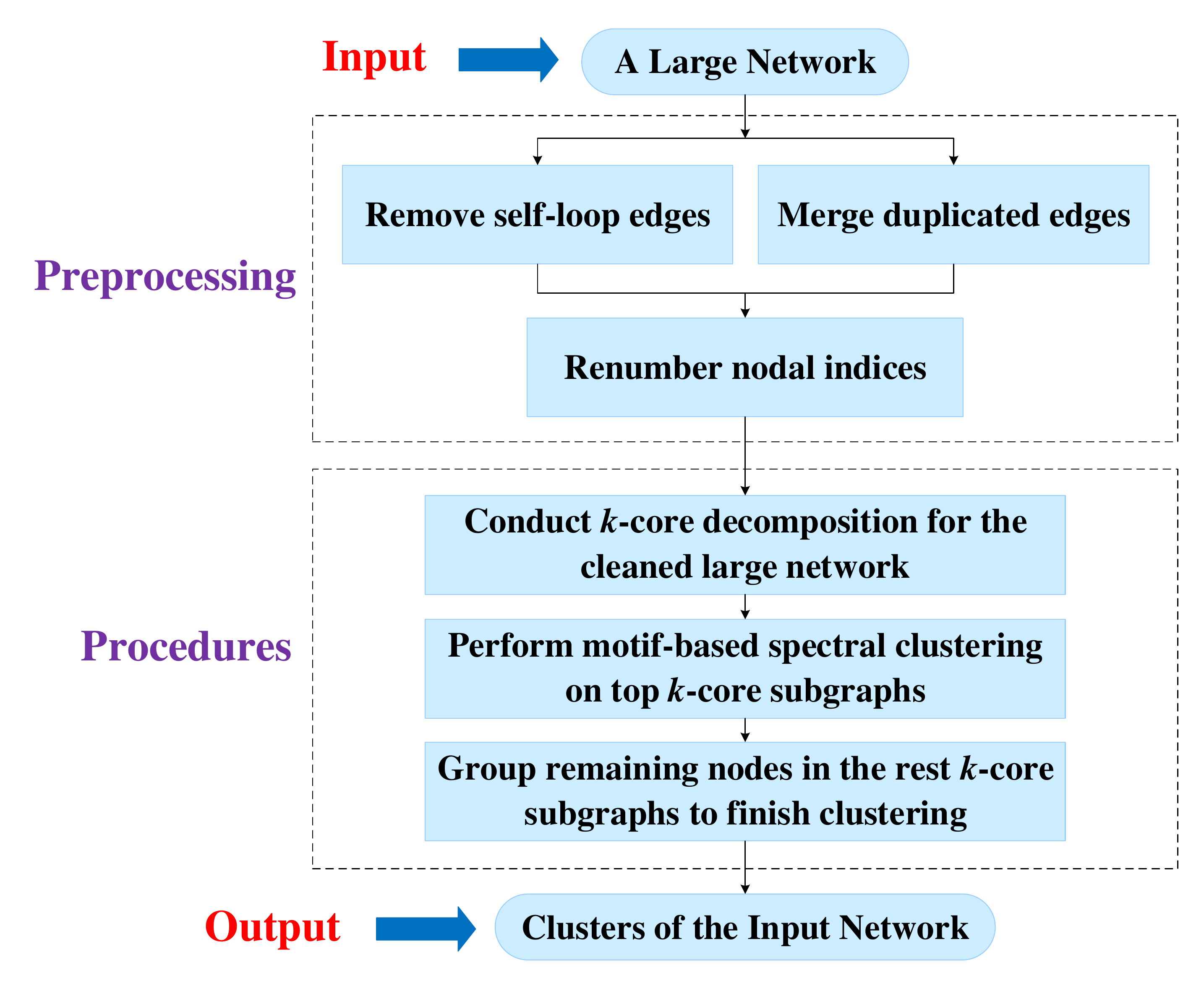}
	\caption{Flowchart of the proposed graph clustering algorithm} 
	\label{Figure5}
\end{figure}

By utilizing the above two ideas, the proposed graph clustering algorithm for working with a large-scale network is composed of three major procedures (Figure \ref{Figure5}):

Procedure 1: Conducting \textit{k}-core decomposition of the large input network to generate \textit{k}-core subgraphs. 

Procedure 2: Performing the motif-based spectral clustering instead of the conventional spectral clustering for the top \textit{k}-core subgraphs.

Procedure 3: Grouping the remaining vertices in the rest (\textit{k}-1)-core subgraphs into previously found clusters to finally obtain the required clusters of the large input network.

It should be noted that, after inputting the original large network and before performing the above three major procedures for clustering, the original large input network is needed to be cleaned up to satisfy the following requirements:

(1)	Self-loop edges need to be removed. That is, if node A and node B of the same edge \textit{e} have the same indices, then the edge \textit{e} is self-looped and thus needs to be removed.
 
 (2) Duplicated edges are merged to be unique. That is, if the first node of an edge, e.g., $ A_{1}$, is exactly the same as the first node of another edge, e.g., $ A_{2}$, while the second nodes of the two edges are also the same, e.g., $ B_{1}$ = $ B_{2}$, then the two directed edges are exactly the same. In this case, these edges are merged to be a unique one, and duplicated ones are removed.
 
(3)	Indices of all nodes are renumbered to be continuous, which starts with 0 and without intervals. This solution of renumbering can be used to reduce the requirements of memory allocation. After finding the clustering, the new indices of all clustered nodes could be recovered to its old indices by exploiting the mapping relationship between the new and old indices.

\subsection{Procedure 1: Conducting \textit{k}-core Decomposition of the Input Large Network}

In recent years, with the increasing interest in the study of complex networks, \textit{k}-core has been widely used as a method for analyzing network topology. The generally accepted concept of \textit{k}-core was first proposed by Seidman \cite{r32}, and he also derived an algorithm called the \textit{k}-core pruning process to obtain the \textit{k}-core of a given network. Due to the simplicity and effectiveness of the \textit{k}-core theory, researchers have applied it to many fields, such as computer science \cite{r33}, social networks \cite{r34}, biology \cite{r35}, and geology study \cite{r36}, etc.

For the graph \textit{G} = (\textit{V}, \textit{E}), where \textit{V} is the set of nodes, and \textit{E} is the set of edges between the nodes. By gradually pruning off less central nodes, leaving a series of increasingly interconnected node sets, we can find the nodes that are critical to the network. When all nodes \textit{i} with degree $ \textit{k}_{i}$ $ \leq$ (\textit{k} - 1) are gradually depleted, the \textit{k}-core is defined as the remaining subset of node \textit{i} with degree $ \textit{k}_{i}$ $ \textgreater$ (\textit{k} - 1). The \textit{k}-shell is defined as a node group that belongs to the \textit{k}-core but does not belong to the (\textit{k} +1)-core.

The following Figure \ref{Figure6} presents a simple illustration of the \textit{k}-core decomposition of a graph. Each node of the graph belongs to 1 core (except for isolated nodes). After gradually cutting off all nodes whose degree is less than 2, the remaining nodes form 2 cores. Then, by further pruning to identify the innermost node set (i.e., 3 cores). Different colors surround different core numbers. It can be seen that as the core level increases, the network scale gradually shrinks. It is observed that the nodes in the highest core are tightly connected, and they are located very close. The number of \textit{k} cores is the topological invariant of the network.

\begin{figure}[!ht]
	\centering
	\includegraphics[width=\textwidth]{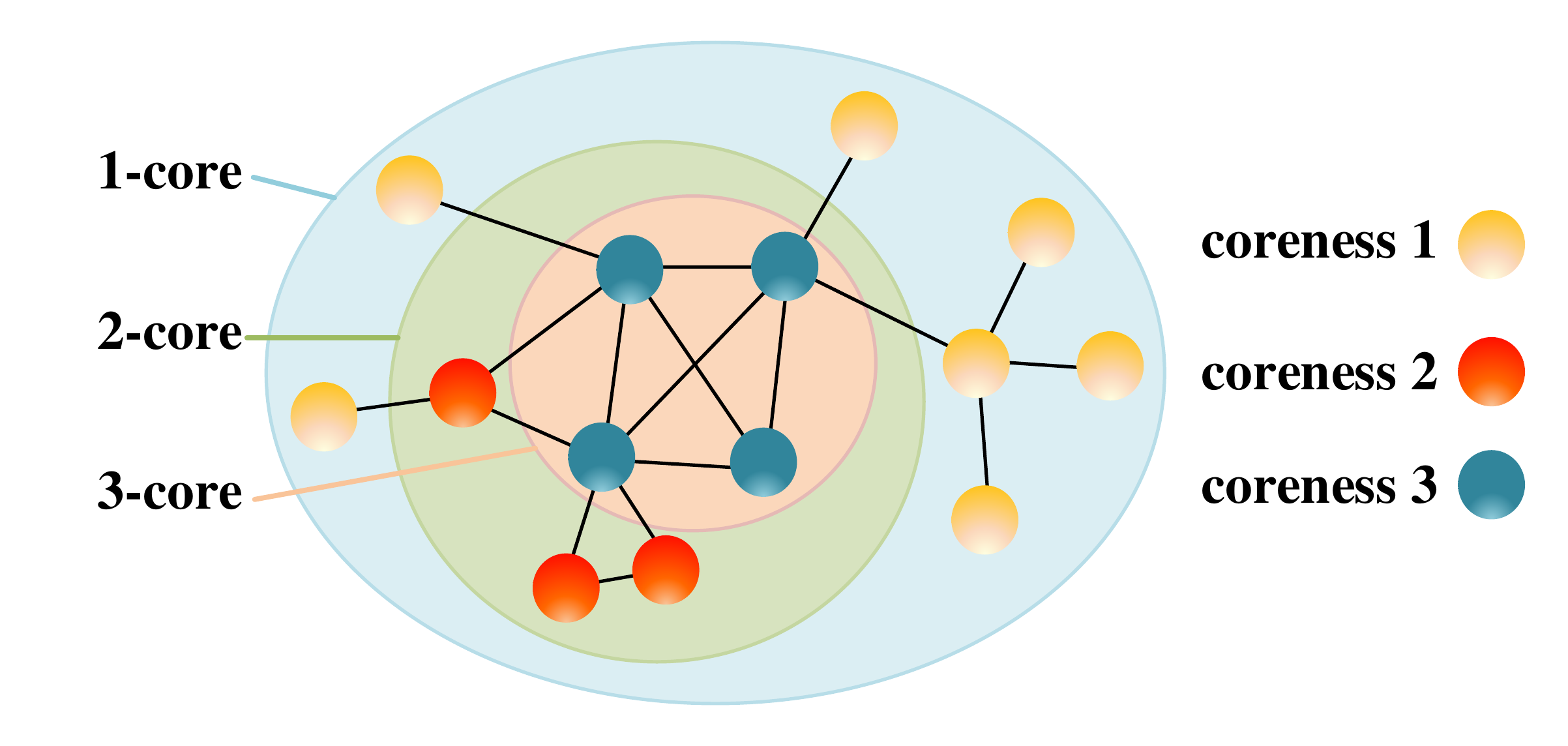}
	\caption{A simple illustration of the \textit{k}-core decomposition} 
	\label{Figure6}
\end{figure}

In our work, we first calculate the coreness of each node by performing the \textit{k}-core decomposition. The efficient \textit{O}(\textit{m}) algorithm for \textit{k}-core decomposition of networks was proposed by Batagelj and Zaversnik \cite{r37}. After the \textit{k}-core decomposition, each node receives its coreness (see Figure \ref{Figure6}).

The reasons why conducting the \textit{k}-core decomposition are as follows. (1) the nodes on the network with large coreness are usually the key members of clusters. That is, the nodes with large coreness commonly are located in the centroid of clusters. Similar to the idea behind the \textit{k}-centroid algorithm, it is better to first cluster these nodes potentially locating in the ‘center’ of centers. (2) the \textit{k}-core decomposition can be used to divide the large input network into a series of components, i.e., the \textit{k}-core subgraphs. Conducting the effective but computationally expensive spectral clustering on the top \textit{k}-core subgraphs, rather than on the entire large network can reduce the computational cost. 

\subsection{Procedure 2: Performing Motif-based Spectral Clustering for the Top \textit{k}-core Subgraphs}

We then perform the motif-based spectral clustering for the specified \textit{k}-core subgraphs. In the subsequent paragraphs, we will first briefly introduce the network motifs and motif-based spectral clustering algorithm.

\begin{figure}[!ht]
	\centering
	\includegraphics[width=\textwidth]{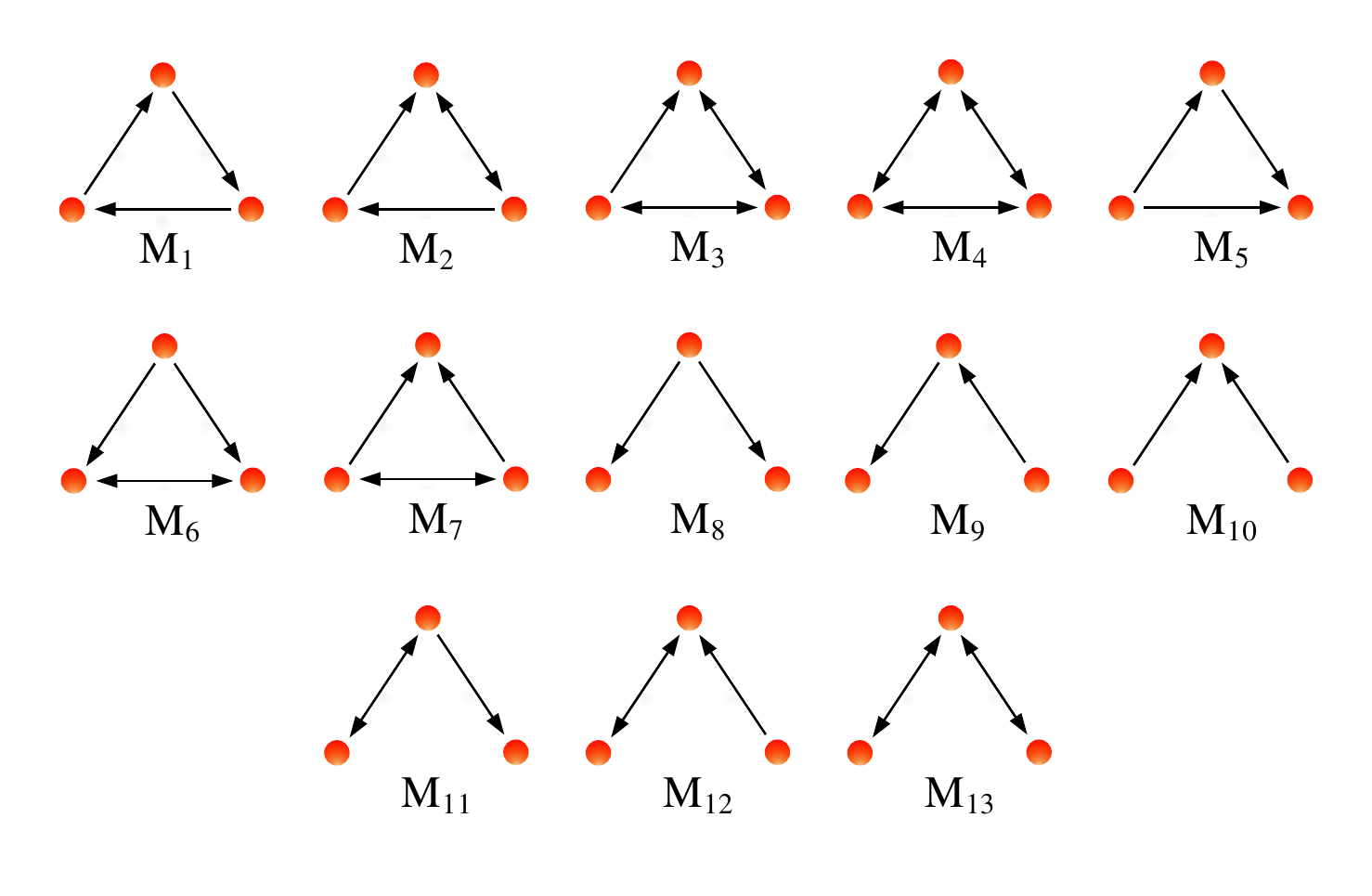}
	\caption{Illustration of all 13 connected three-node directed \cite{r23}} 
	\label{Figure7}
\end{figure}

Networks are not composed only of edges, but these edges form small subgraphs, which we call motifs or graphlets. And these higher-order structures are building blocks of networks. For example, here are all possible directed connected subgraphs on 3 nodes; see Figure \ref{Figure7}.

Recently, one of the novel ideas is to utilize the small subgraphs (motifs/graphlets) to enhance the efficiency and scalability of finding clusters (communities) on large networks. Among those research works, the most famous one is the motif-based spectral clustering algorithm proposed by Austin Benson, David Gleich, and Jure Leskovec \cite{r23}. 

The most essential and interesting idea behind the aforementioned motif-based spectral clustering algorithm is to build the adjacency and Laplacian matrices based on the found network motifs, rather than based on the network edges where it is usually conducted in the conventional spectral clusterings. In other words, in the conventional spectral clustering algorithm, both the adjacency and Laplacian matrices are built on the basis of the topological relationships representing by edges, while in contrast in the motif-based spectral clustering algorithm, both the adjacency and Laplacian matrices are assembled on the basis of the topological relationships representing by motifs. In short, nodes on the networks are clustered based on motifs instead of edges. 

The process of the motif-based spectral clustering algorithm is quite similar to that of the conventional spectral clustering algorithm. 

(1) Given a type of motif M (e.g., M6) and a graph \textit{G}, find the distribution of the motifs;

(2) Assemble the weighted adjacency matrix \textit{W} first and then the Laplacian matrix \textit{L} based on the distribution of motifs;

(3) Calculate the eigenvalues and corresponding eigenvectors of the Laplacian matrix \textit{L};

(4) Conduct \textit{k}-means clustering on the aforementioned eigenvectors to obtain the desired clusters of nodes.

In our work, we employ the aforementioned motif-based spectral clustering algorithm to cluster the local sets of nodes on the specified \textit{k}-core subgraphs. Moreover, the motif-based clustering for the top \textit{k}-core subgraphs is performed by modifying and revising the related source code introduced in the reference \cite{r23}. Note that, in this procedure, one of the critical issues is the determination of the \textit{k} value when selecting the \textit{k}-core subgraphs for performing the motif-based spectral clustering. The values of \textit{k} could have a strong impact on the efficiency and accuracy of the proposed graph clustering algorithm. For example, in general, large \textit{k} value will increase the accuracy of finding clusters but will decrease the efficiency, while in contrast, small \textit{k} value will decrease the accuracy but increase efficiency. We will discuss this issue and give our suggestions later in the section of the discussion. 

\subsection{Procedure 3: Grouping the Remaining Vertices in the Rest \textit{k}-core subgraphs into Previously Found Clusters}

In this procedure, we group the remaining vertices in the rest \textit{k}-core subgraphs into those previously found clusters and obtain the required clusters of the large input network. 

In the above procedure, the motif-based spectral clustering is performed on the top \textit{k}-core subgraphs. After the clustering, each node locating on the top \textit{k}-core subgraphs will have its own label of clusters, where the label indicates which cluster it belongs to. However, those nodes outside the top \textit{k}-core subgraphs do not have their labels of clusters. These unlabeled nodes need to be labeled according to the existing clusters. 

Similar to the basic idea behind the “label propagation” algorithm \cite{r38,r39}, in this procedure, the labels of those unlabeled nodes are assigned according to the labels of their neighbors. More specifically, if a node $ A$ has 10 neighboring nodes, and among those 10 neighbors, 5 nodes have been labeled and belong to a cluster denoted as $ C_{1}$, then the node $ A$ will also be assigned to the cluster $ C_{1}$.

However, when assigning the nodal label according to the labels of its neighbors, there is a probable problem: little or even no neighbors of the target node have been assigned labels. In this case, the target node cannot be assigned to a label according to information of its neighboring nodes. Thus, in this situation, a new cluster is created, and the index of the new cluster will be assigned to this node. 

In summary, the process of assigning the label of an unlabeled node $ A$ according to the clustering information of its neighbors is as follows.

(1)	Loop over all neighbors of the node $ A$, and accumulate the numbers of neighbors locating in the same cluster;

(2)	Calculate the percentage of the number of neighbors locating in the same cluster over the total number of neighbors;

(3)	Sort the percentages in descending order;

(4)	Check whether the largest percentage exceeds a specified threshold, e,g., 50$\%$, if exceeding, then assign the target node $ A$ to the cluster that most of its neighbors locate in; otherwise, allocate a new cluster and assign the target node $ A$ to the new cluster. 

It should be clearly noted that, the recovery of the node label is conducted by “shell-by-shell”, which is different from the recovering processed introduced in the references \cite{r21,r22}. More specifically, after performing the motif-based spectral clustering algorithm on the specified \textit{k}-core subgraph, each node on the \textit{k}-core subgraph will have a label indicating its ID of cluster. When grouping the remaining nodes on the rest (\textit{k}-1)-core subgraph, we first recover those nodes on the (\textit{k}-1)-shell, and then the (\textit{k}-2)-shell, (\textit{k}-3)-shell, and so on. The above recovering is carried out “shell-by-shell” until all the remaining nodes on the rest (\textit{k}-1)-core subgraph are grouped into previously found clusters.

\subsection{Implementation Details of the Proposed Graph Clustering Algorithm}

The proposed graph clustering algorithm is implemented in the Julia language. More details about the Julia implementation of the major procedures of the proposed graph clustering algorithm are described as follows. 

First, when implementing the procedure of \textit{k}-core decomposition, we exploit the related functions provided by the package \texttt{LightGraphs.jl}. The functions \texttt{k\_core()}can be used to find nodes with coreness that is greater than or equal to \textit{k}, while the function \texttt{k\_crust()} is capable of finding nodes with coreness that is less than \textit{k}. In addition, function \texttt{core\_number()} is used to calculate the core number of each node. Although most steps of the \textit{k}-core decomposition have been implemented by the package \texttt{LightGraphs.jl}, the package lacks the features of forming a \textit{k}-core subgraph. Thus, function \texttt{g\_core4()} is specifically implemented by ourselves. The adjacency matrix of the \textit{k}-core subgraph is obtained when the normal adjacency matrix minus the adjacency matrix of the core subgraph. In this step, the key is to find the deleted nodes and edges, the function \texttt{k\_crust()} is used to find the deleted nodes. Moreover, functions \texttt{outneighbors()} and \texttt{inneighbors()} are used to find the out-neighbors and in-neighbors of the deleted node. The above results will be used to perform the local motif-based spectral clustering.

Second, when implementing the local motif-based spectral clustering, the motif adjacency matrix is formed according to the selected motif type, and then the core subgraph is clustered according to the traditional spectral clustering to get the cluster to which each node belongs. To form the motif-based adjacency matrix, the operation min.(A, A$ '$)\textit{ }( \textbf{.} operation can realize the operation of the array by element)\textit{ }is used to obtain the all bidirectional edges, and then the functions \texttt{M1()} $\sim$ \texttt{M7()} are used to implement the formation of the motif adjacency matrix. Finally, to realize the function of traditional spectral clustering, we integrate the formation of the Laplacian matrix, the normalization of the Laplacian matrix, and the calculation of eigenvector into function \texttt{spectral\_embedding()}. Moreover, function \texttt{eigs()} in package \texttt{Arpack.jl} is employed to calculate the eigenvectors.

Third, when implementing the label recovery to group the remaining nodes on the rest (\textit{k}-1)-core subgraph to previously found clusters, function \texttt{labelrecover()} is written to implement this step. The labels of filtered nodes are recovered from the nodes with higher \textit{k} value to the nodes with lower \textit{k }value. Details are as follows.

(1) All labels of filtered nodes are set to 0. Function \texttt{initial()} is written to implement this step.

(2) The coreness of all filtered nodes are found. The \texttt{k\_crust()} function can be used to implement this step.

(3) The remaining nodes are sorted in descending order according to their coreness. In the step, for a more convenient index, an\textit{ }array with type \texttt{Pair} is created to connect the ID of each deleted node to its coreness. On the other hand, the sorting function \texttt{sort()} in the Julia language can easily sort the whole Pair array according to the value or key of the \texttt{Pair} array element. In addition, to save running time for handling large datasets, the quick sort algorithm \texttt{QuickSort} is selected.

(4) From the node with the maximum coreness, the frequencies of labels of the nodes which have been assigned label in the nodal (i.e., the maximum coreness) neighbors are counted. In this step, the function \texttt{all\_Neighbors()} can be used to easily obtain all the neighbors of a node. However, we only count the frequencies of labels of the nodes which have been assigned a label. Thus, the nodes with label 0 are not counted.

(5) The neighbors of the node with the maximum coreness are sorted according to the frequencies of labels. In this step, a similar operation to step 3 is used, that is, sorting a \texttt{Pair} array.

(6) The label with the largest frequency is the label of the filtered node.

\section{Results}
\label{sec4}

To evaluate the performance of the proposed graph clustering algorithms, we use the conventional spectral clustering and the motif-based spectral clustering algorithms as the baseline and compare both the accuracy and efficiency of our algorithm with them for 18 groups of real-world datasets. All the testing datasets can be download from the website of SNAP (\url{http://snap.stanford.edu}). And all the source code of the implementations of our algorithm is available associated with this paper. 

\subsection{Experimental Setup}

All experimental tests are conducted on a powerful workstation computer. Specifications and configurations of the employed experimental setup are listed in Table \ref{tab1}.

\begin{table}[H]
	\caption{Specifications of the workstation computer for testing the proposed algorithm}
	\begin{center}
		 \begin{tabular}{|p{1.5in}|p{1.7in}|} \hline 
		 	\textbf{Specifications} & \textbf{Details} \\ \hline 
		 	\textbf{CPU} & Intel Xeon Gold 5118 CPU \\ \hline 
		 	\textbf{CPU Frequency (GHz)} & 2.30 \\ \hline 
		 	\textbf{CPU RAM (GB)} & 128 \\ \hline 
		 	\textbf{CPU Core} & 48 \\ \hline 
		 	\textbf{OS} & Windows 10 Professional \\ \hline 
		 	\textbf{Visual Studio} & VS 2015 Community \\ \hline 
		 	\textbf{Python} & Version 3.7 \\ \hline 
		 	\textbf{Julia} & Version 1.4.1 \\ \hline 
		 \end{tabular}
		 \label{tab1}
	\end{center}
\end{table}

\subsection{Real-world Datasets for Testing}

The large networks for the benchmark tests are download from the Stanford Network Analysis Platform (SNAP). Details of those large networks are listed in Table \ref{tab2}.

\begin{table}[H]
	\small
	\caption{Datasets used for experimental tests}
	\begin{center}
		\begin{tabular}{|p{1.1in}|p{0.56in}|p{0.56in}|p{2.65in}|} \hline 
			\textbf{Name} & \textbf{Nodes} & \textbf{Edges} & \textbf{Description} \\ \hline 
			reachability & 456 & 71959 & City reachability through air travel \\ \hline 
			web-Stanford & 281903 & 2312497 & Web graph of Stanford.edu \\ \hline 
			ca-HepTh & 9877 & 25998 & Collaboration network of Arxiv High Energy Physics Theory \\ \hline 
			ca-CondMat & 23133 & 93497 & Collaboration network of Arxiv Condensed Matter \\ \hline 
			musae-facebook & 22470 & 171002 & Facebook page-page network with page names \\ \hline 
			Facebook\_combined & 4039 & 88234 & Participants using this Facebook app \\ \hline 
			bitcoinalpha & 3783 & 24186 & Network of people who trade using Bitcoin on a platform called Bitcoin Alpha \\ \hline 
			lastfm\_asia & 7624 & 27806 & A social network of LastFM users \\ \hline 
			deezer\_europe & 28281 & 92752 & A social network of Deezer users \\ \hline 
			wiki-Talk & 2394385 & 5021410 & Wikipedia talk (communication) network \\ \hline 
			soc-Pokec & 1632803 & 30622564 & Pokec online social network \\ \hline 
			amazon0505 & 410236 & 3356824 & Amazon product co-purchaisng network from May 05 2003 \\ \hline 
			amazon0302 & 262111 & 1234877 & Amazon product co-purchaisng network from March 02 2003 \\ \hline 
			web-BerkStan & 685230 & 7600595 & Berkely-Stanford web graph from 2002 \\ \hline 
			wiki-topcats & 1791489 & 28511807 & Hyperlink network of Wikipedia \\ \hline 
			web-Google & 875713 & 5105039 & Webgraph from the Google programming contest, 2002 \\ \hline 
			wiki-talk & 1140149 & 3309592 & Wikipedia users editing each other's Talk page \\ \hline 
			sx-stackoverflow & 2464606 & 16266395 & Interactions on the stack exchange web site Stack Overflow \\ \hline 
		\end{tabular}
		\label{tab2}
	\end{center}
\end{table}

\subsection{Experimental Results and Analysis}

\subsubsection{Accuracy of the Proposed Graph Clustering Algorithm}

We first investigate the accuracy of the proposed graph clustering algorithm, KCoreMotif, using 8 datasets based on the commonly used measures of graph clustering quality, i.e., the Modularity (Q) \cite{r40}; see Table \ref{tab3}.

\begin{table}[H]
	\small
	\caption{Comparison of the accuracy (modularity) of three graph clustering algorithm}
	\begin{center}
		\begin{tabular}{|p{1.15in}|p{0.75in}|p{0.71in}|p{0.995in}|p{0.995in}|} \hline 
			\textbf{Dataset} & Conventional & Motif-based & KCoreMotif-50\% & KCoreMotif-80\% \\ \hline 
			CA-CondMat & 0.2092 & 0.0019 & 0.0144 & 0.0872 \\ \hline 
			CA-HepTh & 0.0689 & 0.0008 & 0.0520 & 0.1236 \\ \hline 
			reachability & 0.1117 & 0.0270 & 0.0074 & N/A \\ \hline 
			musae-facebook & 0.2292 & 0.0018 & 0.0481 & 0.1337 \\ \hline 
			Facebook\_combined & 0.5598 & 0.0019 & 0.0666 & 0.1861 \\ \hline 
			bitcoinalpha & 0.0093 & 0.0028 & 0.0071 & 0.1046 \\ \hline 
			lastfm\_asia & 0.6051 & 0.0005 & 0.0752 & 0.1642 \\ \hline 
			deezer\_europe & 0.3937 & 0.0002 & 0.0853 & 0.2058 \\ \hline 
		\end{tabular}
		\label{tab3}
	\end{center}
\end{table}

Note that, in the proposed graph clustering algorithm, the \textit{k}-core decomposition is first performed, and then the motif-based spectral clustering is conducted on the top \textit{k}-core subgraphs. That is, by exploiting the \textit{k}-core decomposition, a part of nodes and edges (e.g., 50\% or 80\% of nodes) are filtered, and the remaining nodes and edges (e.g., 50\% or 20\% of nodes) are used further to carry out the spectral clustering. 

Here in this work, we also investigate the accuracy of the proposed graph clustering algorithm in the two cases where 50\% and 80\% of nodes are filtered, respectively. And we compare the accuracy with those of the conventional spectral clustering and the famous motif-based spectral clustering proposed by Jure Leskovec \cite{r23}.

The aforementioned comparative results listed in Table \ref{tab3} indicate that: 

(1) In general, the conventional spectral clustering can achieve the best accuracy. 

(2) The proposed graph clustering algorithm, KCoreMotif, can achieve a little better accuracy than the motif-based clustering algorithm. 

(3) The accuracy of clustering in the case where approximately 50\% nodes are filtered is worse than that where 80\% nodes are filtered. We will specifically explain this in the section of the discussion. 

\subsubsection{Efficiency of the Proposed Graph Clustering}

In this subsection, we will evaluate the computational efficiency of the proposed graph clustering algorithm by comparing it with (1) the conventional spectral clustering algorithm and (2) the motif-based spectral clustering algorithm. 

When evaluating the performance of the proposed graph clustering algorithm, we specifically investigate the distribution of the nodal coreness of the testing networks. We find that, similar to the classic distribution of degree centrality; most testing networks are with (1) the normal distribution of the nodal coreness or (2) the power-law distribution of the nodal coreness. Thus, in this work, we evaluate the clustering efficiency for the above two types of networks.

\textbf{(1) Efficiency for networks with the normal distributions of coreness}

\begin{figure}[H] 
	\centering
	\subfigure[26w-Amazon0302]{
		\label{Figure8a}       
		\includegraphics[width=0.435\textwidth]{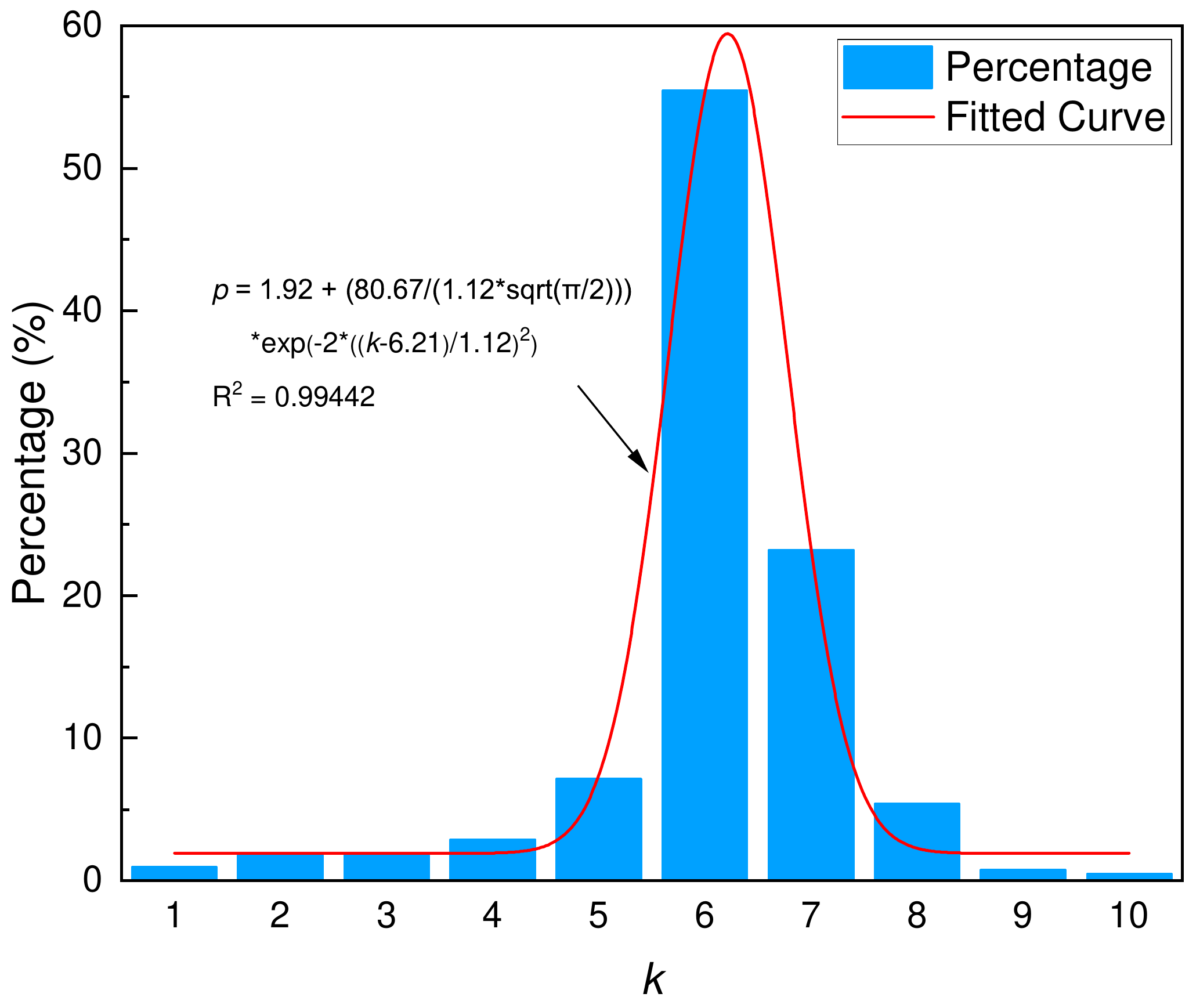}
	}
	\subfigure[28w-web-Stanford]{
		\label{Figure8b}       
		\includegraphics[width=0.435\textwidth]{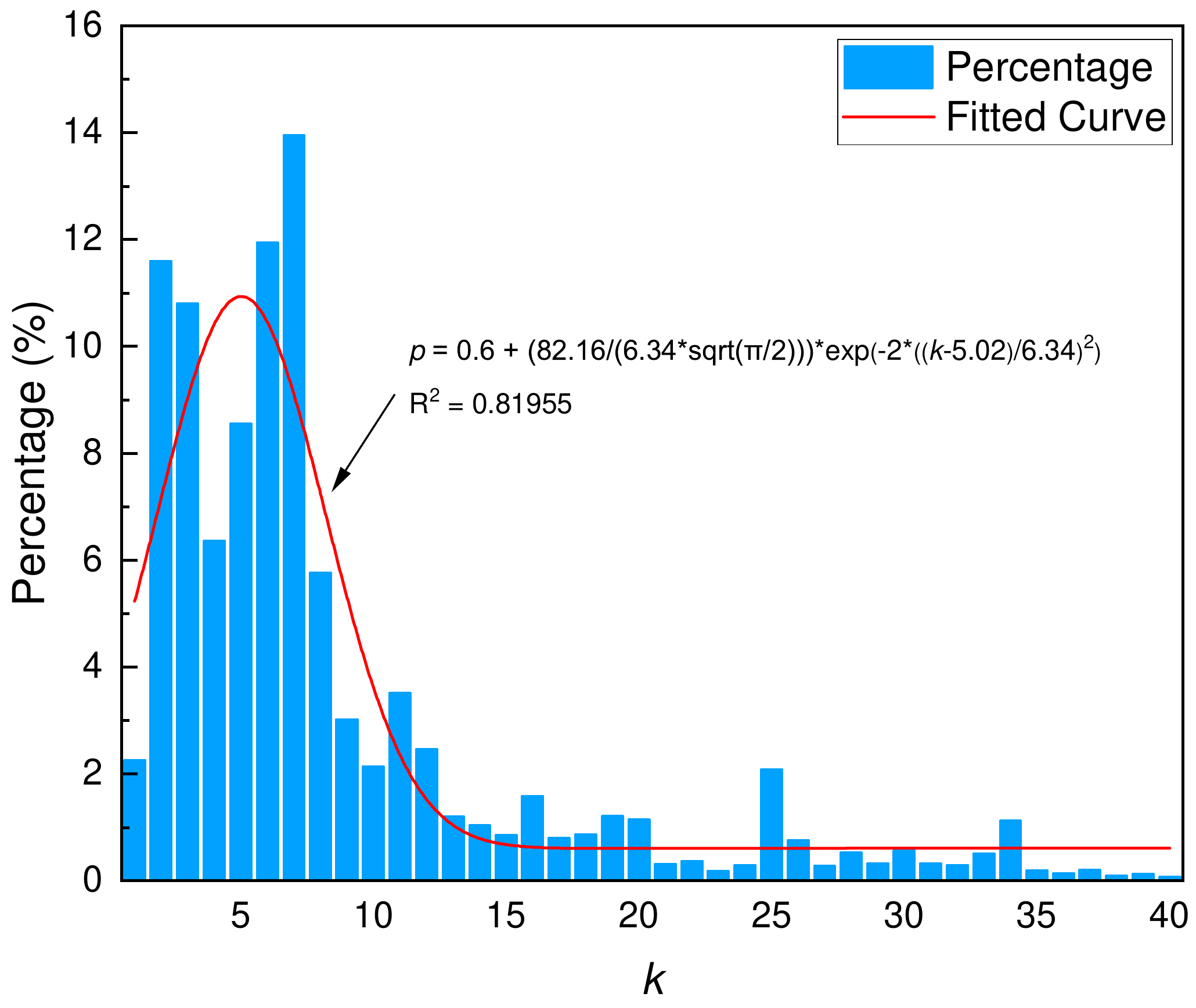}
	}
	\subfigure[41w-Amazon0505]{
		\label{Figure8c}       
		\includegraphics[width=0.435\textwidth]{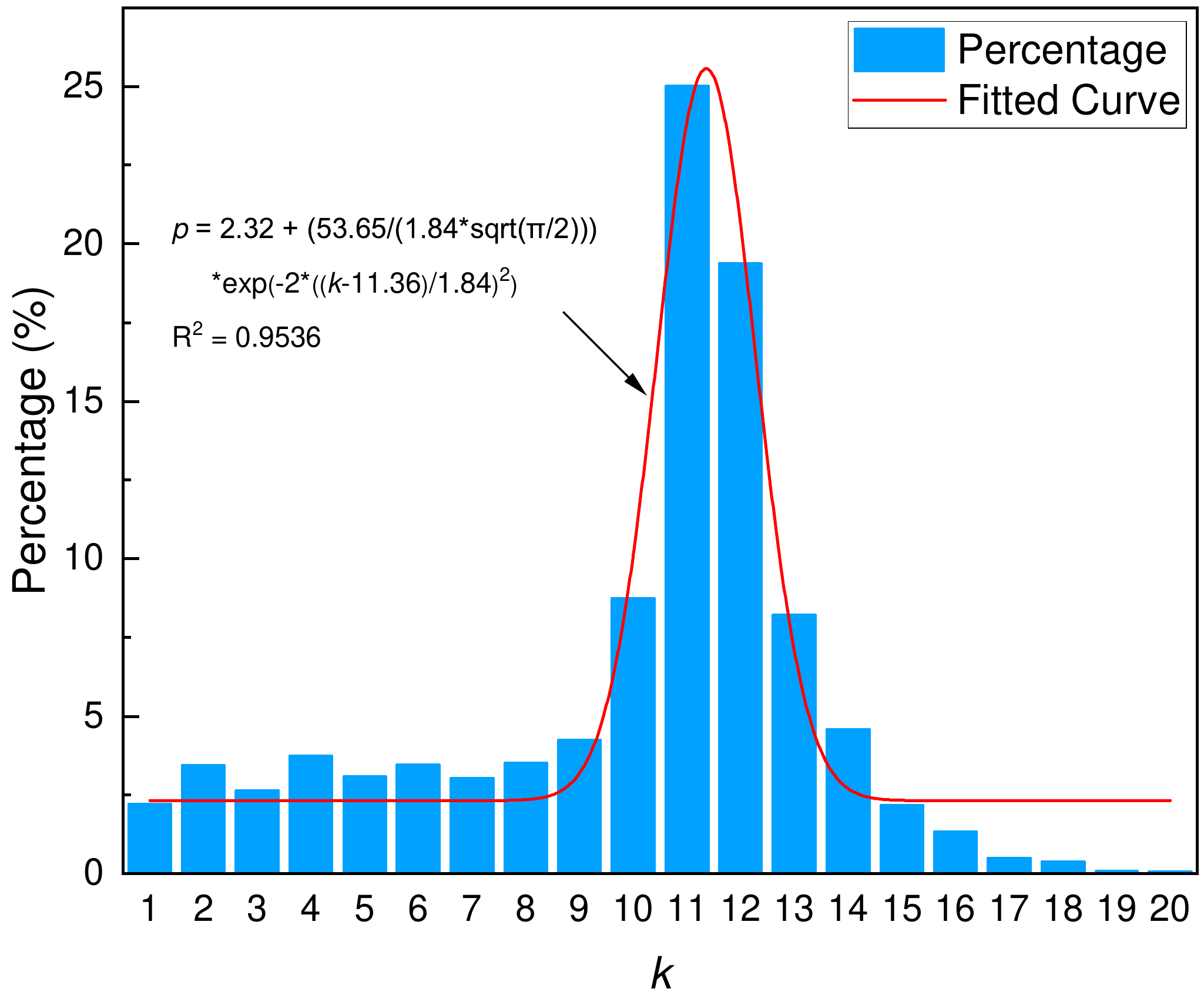}
	}
	\subfigure[68w-web-BerkStan]{
		\label{Figure8d}       
		\includegraphics[width=0.435\textwidth]{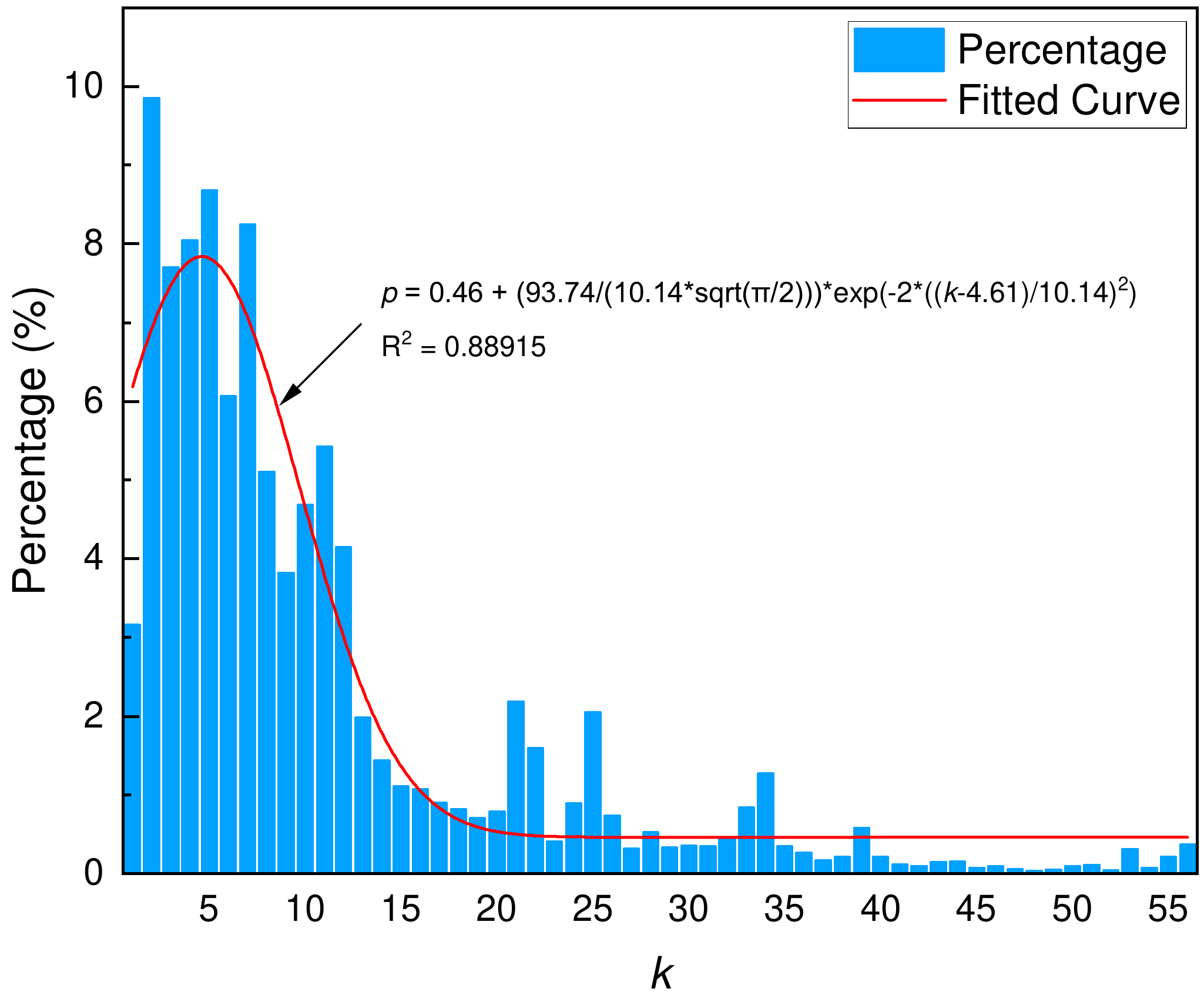}
	}
	\subfigure[170w-wiki-topcats]{
		\label{Figure8e}       
		\includegraphics[width=0.435\textwidth]{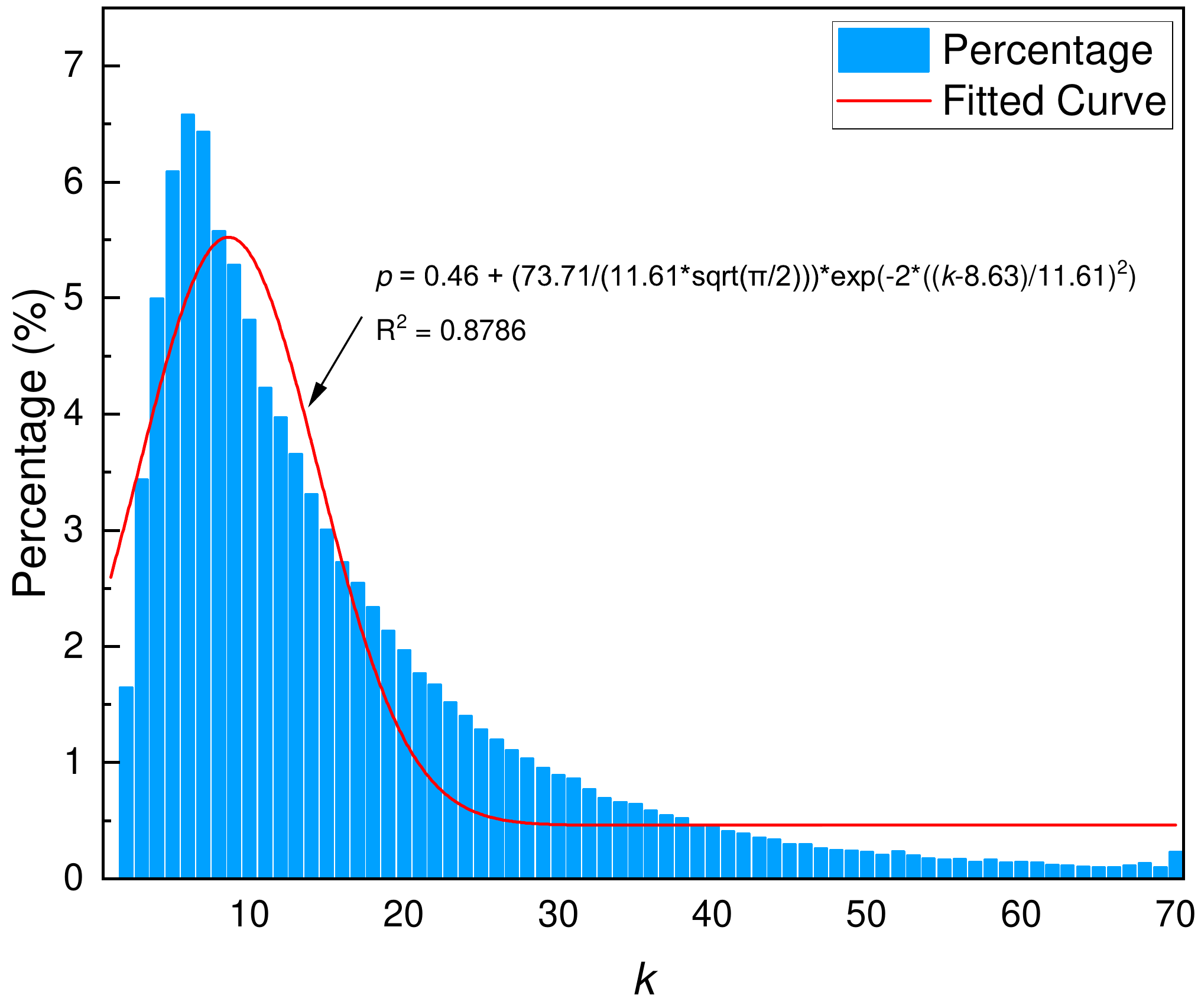}
	}
	\caption{Networks with the normal distributions of coreness}
	\label{Figure8}       
\end{figure}

We use five networks with the normal distribution of nodal coreness for testing; see Figure \ref{Figure8}; The running time of the three algorithms for 5 groups of datasets is comparatively listed in Table \ref{tab4}. 

The comparative results indicate that: (a) the conventional spectral clustering is the most computationally expensive, and cannot work with large-scale networks; (b) the motif-based spectral clustering algorithm is much faster than the conventional spectral clustering algorithm, and can work with large-scale networks; (c) the proposed graph clustering algorithm, KCoreMotif, is faster than the motif-based spectral clustering; and in the best cases, it can achieve the speedups of approximately 3x over the motif-based spectral clustering.

Note that, the values of \textit{k} in the \textit{k}-core decomposition strongly affect the clustering efficiency. We will evaluate the impact the values of \textit{k} on both the accuracy and efficiency of clustering, and suggest the selection of \textit{k} values in practices; please refer to Subsection 5.1.

\begin{table}[H]
	\small
	\caption{Comparision of the efficiency of three algorithms for 5 networks with the normal distribution of coreness}
	\begin{center}
		\begin{tabular}{|p{0.76in}|p{0.55in}|p{0.45in}|p{0.35in}|p{0.8in}|p{0.45in}|p{0.36in}|p{0.6in}|p{0.35in}|} \hline 
			\textbf{Dataset} & \multicolumn{3}{|p{1.4in}|}{\textbf{After cleaning up}} & {\textbf{Conventional}}& \textbf{Motif-based} & \multicolumn{3}{|p{1.5in}|}{\textbf{Our KCoreMotif}} \\ \hline 
			& Edges & Nodes & E / N & Time & Time & \textit{k}-core & Remaining & Time \\ \hline 
			\textbf{Amazon0302} & 1234877 & 262111 & 46.73 & 419.40 & 31.08 & 7 & 29.89\% & 10.14 \\ \hline 
			\textbf{web-Stanford} & 2312497 & 281903 & 82.03 & 620.18 & 63.22 & 9 & 29.92\% & 18.29\newline  \\ \hline 
			\textbf{Amazon0505} & 3356824 & 410236 & 8.18 & 1109.39 & 34.24 & 12 & 52.04\% & 31.00 \\ \hline 
			\textbf{web-BerkStan} & 7600595 & 685230 & 110.92 & N/A & 122.83 & 7 & 57.43\%\newline  & 114.29 \\ \hline 
			\textbf{wiki-topcats} & 28508140 & 1791489 & 15.91 & N/A & 292.96 & 40 & 6.99\%\newline  & 102.03 \\ \hline 
		\end{tabular}
		\label{tab4}
	\end{center}
\end{table}

\textbf{(2) Networks with the power-law distribution of coreness}

\begin{figure}[H] 
	\centering
	\subfigure[87w-webGoogle]{
		\label{Figure9a}       
		\includegraphics[width=0.435\textwidth]{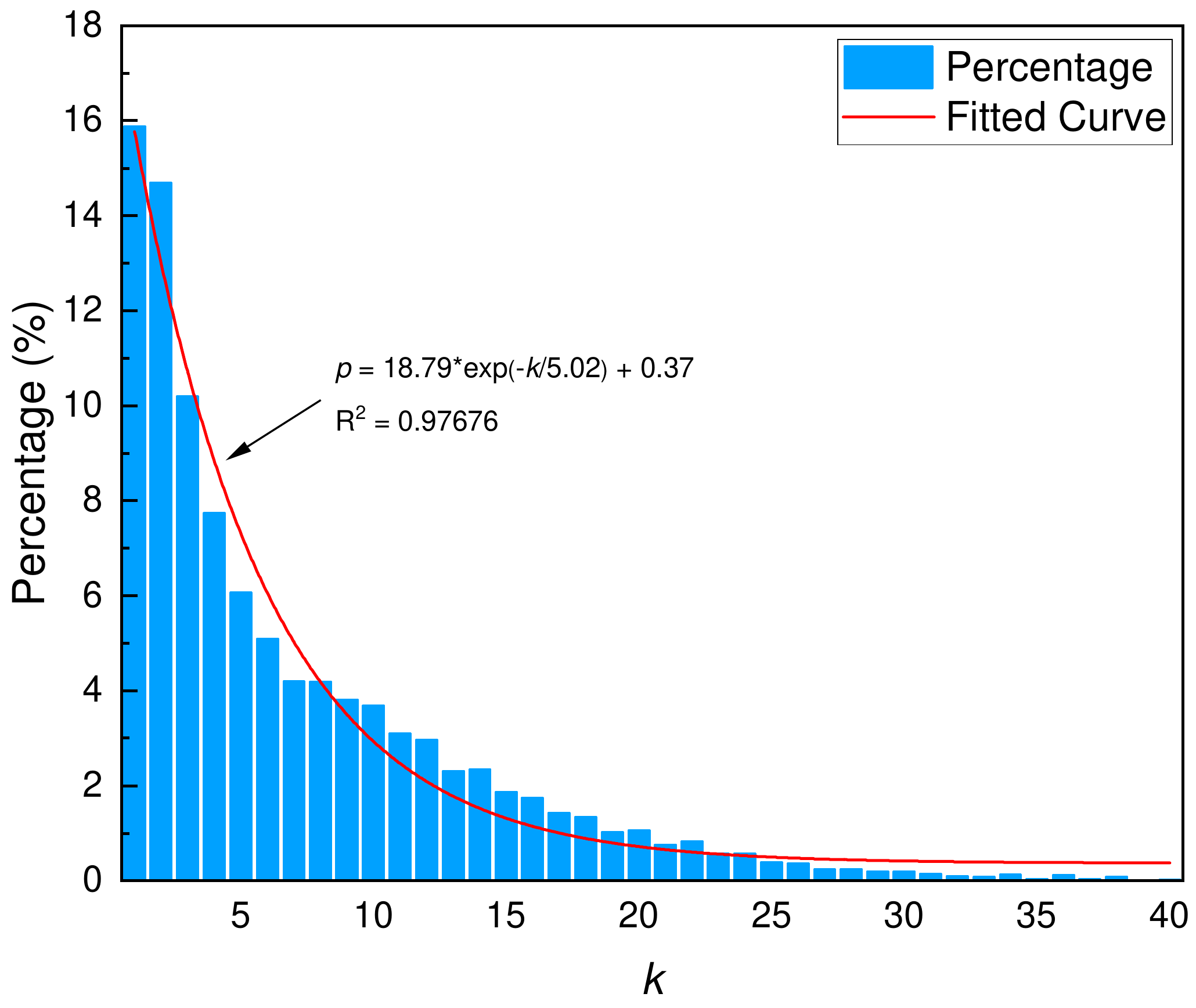}
	}
	\subfigure[110w-wikitalk]{
		\label{Figure9b}       
		\includegraphics[width=0.435\textwidth]{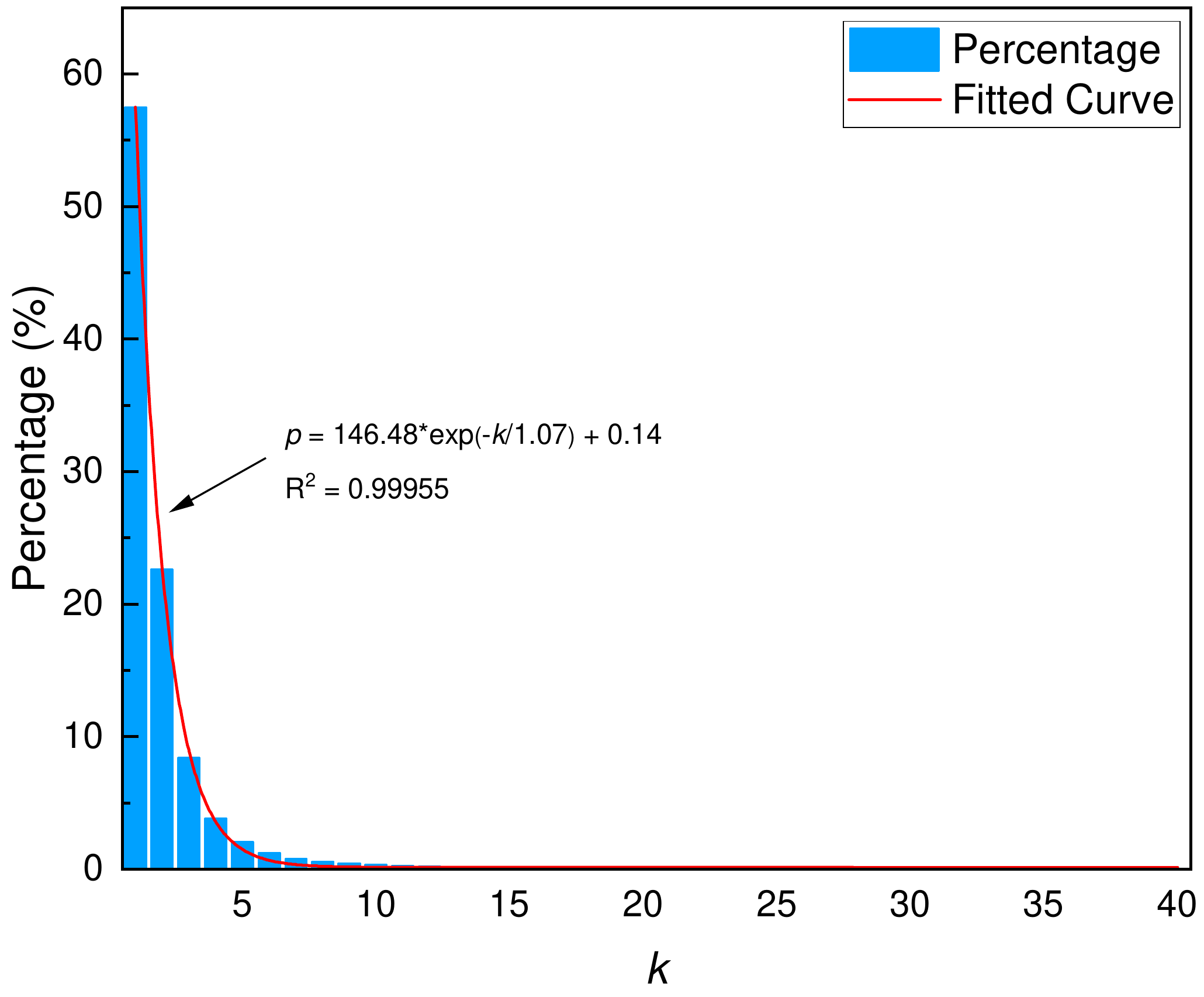}
	}
	\subfigure[160w-soc-pokectxt]{
		\label{Figure9c}       
		\includegraphics[width=0.435\textwidth]{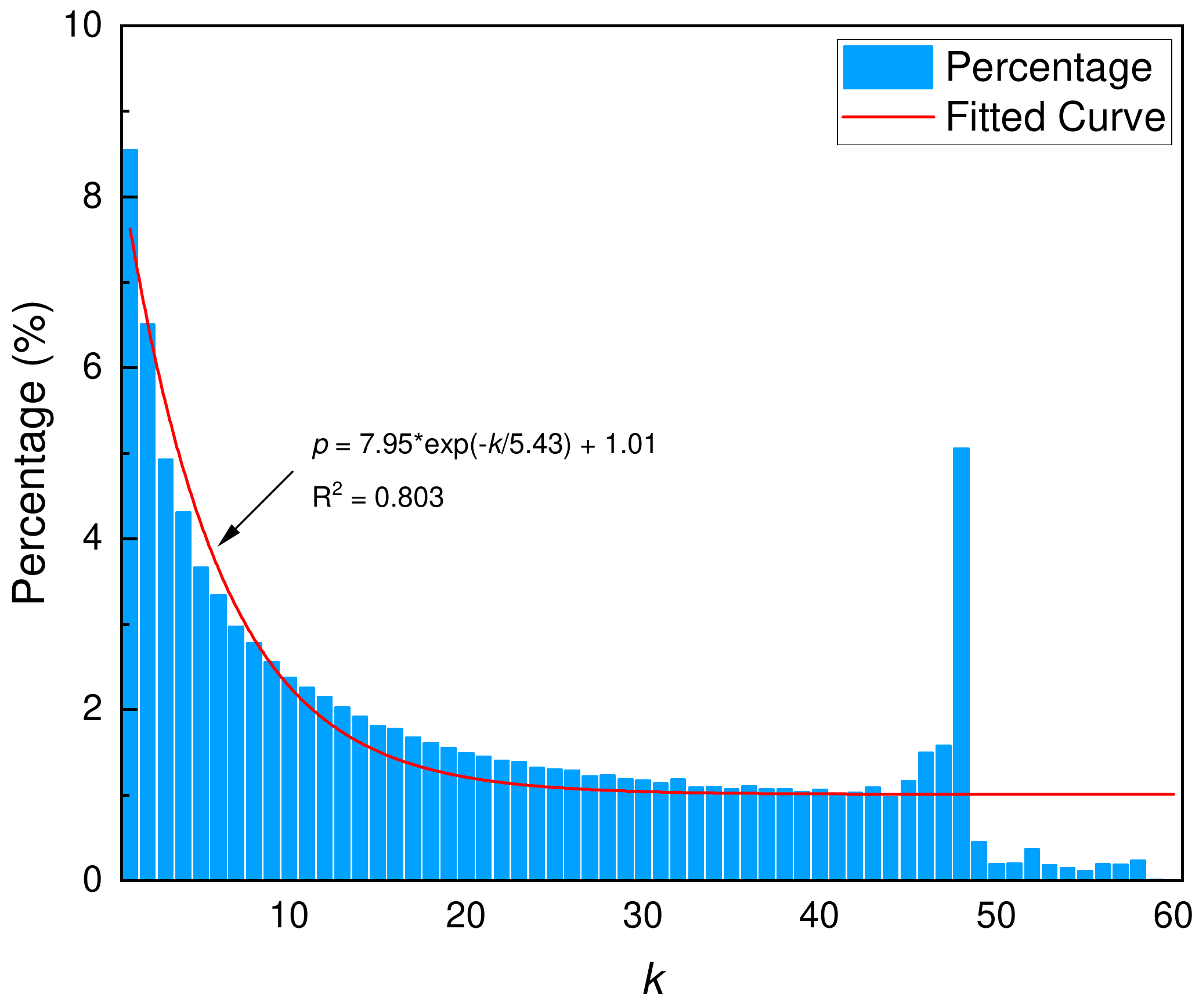}
	}
	\subfigure[230w-WIKItalk]{
		\label{Figure9d}       
		\includegraphics[width=0.435\textwidth]{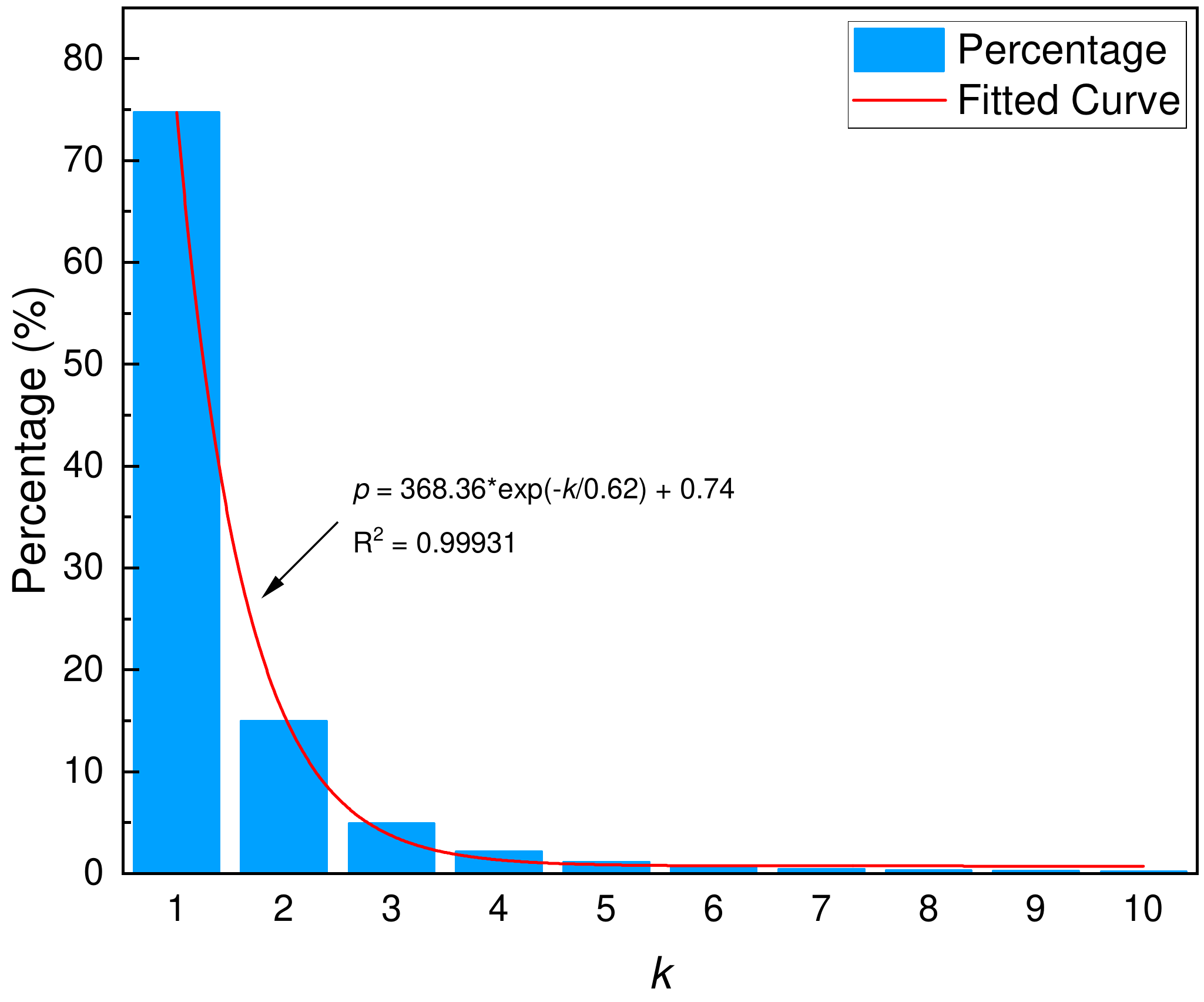}
	}
	\subfigure[250w-sx-stackoverflow]{
		\label{Figure9e}       
		\includegraphics[width=0.435\textwidth]{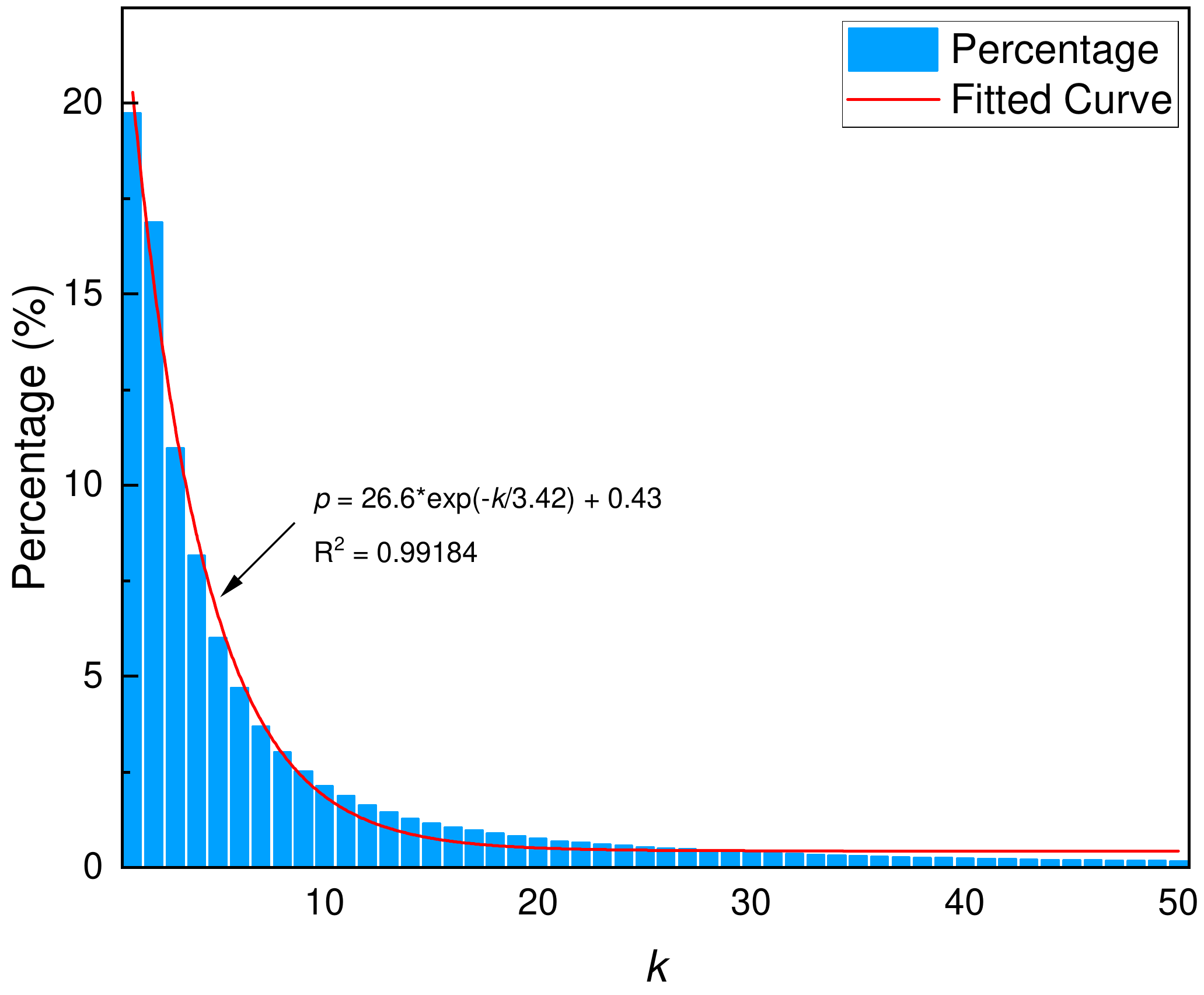}
	}
	\caption{Networks with the power-law distributions of coreness}
	\label{Figure9}       
\end{figure}

Similarly, we use five networks with the power-law distribution of nodal coreness for testing; see Figure \ref{Figure9}; The running time of the three algorithms for 5 groups of datasets is comparatively listed in Table \ref{tab5}.

\begin{table}[H]
	\small
	\caption{Comparision of the efficiency of three algorithms for 5 networks with the power-law distribution of coreness}
	\begin{center}
		\begin{tabular}{|p{0.78in}|p{0.55in}|p{0.5in}|p{0.33in}|p{0.8in}|p{0.4in}|p{0.36in}|p{0.6in}|p{0.35in}|} \hline 
			\textbf{Dataset} & \multicolumn{3}{|p{1.4in}|}{\textbf{After cleaning up}} & \textbf{Conventional} & \textbf{Motif-based} & \multicolumn{3}{|p{1.5in}|}{\textbf{Our KCoreMotif}} \\ \hline 
			& Edges & Nodes & E / N & Time & Time & \textit{k}-core & Remaining & Time \\ \hline 
			webGoogle & 5105039 & 875713 & 5.83 & N/A & 262.70 & 28 & 1.47\% & 34.78 \\ \hline 
			wiki-talk-temporal & 3130741 & 1094017 & 2.86 & N/A & 42.10 & 10 & 3.30\%\newline  & 21.50 \\ \hline 
			soc-Pokec & 30622563 & 1632803 & 18.75 & N/A & 243.51 & 40 & 16.88\% & 127.67 \\ \hline 
			wiki-Talk & 5021409 & 2394384 & 2.10 & N/A & 84.16 & 10 & 2.18\% & 52.95 \\ \hline 
			sx-stackoverflow & 34875683 & 2584163 & 13.49 & N/A & 692.53 & 80 & 3.40\%\newline  & 208.01 \\ \hline 
		\end{tabular}
		\label{tab5}
	\end{center}
\end{table}

The comparative results indicate that: (a) the conventional spectral clustering cannot work with these 5 large-scale networks; (b) the motif-based spectral clustering algorithm can work with these 5 large-scale networks; (c) the proposed graph clustering algorithm, KCoreMotif, is faster than the motif-based spectral clustering; and in the best cases, it can achieve the speedups of approximately 2x $\mathrm{\sim}$ 3x over the motif-based spectral clustering. 

For  networks with the power-law distribution of nodal coreness, in the section of  discussion, we will evaluate the impact the values of \textit{k} on both the accuracy and efficiency of clustering, and suggest the selection of \textit{k} values(see Subsection 5.1).

\subsubsection{Workload of Procedures in the Proposed Graph Clustering Algorithm}

As introduced several times, there are three major procedures of the proposed graph clustering algorithm (see more details of those major procedures in Subsection 3.2). To identify the potential bottleneck of the efficiency of the proposed algorithm, we specifically analyze the workload of the procedures of the proposed graph clustering algorithm. In other words, we record and compare the computational cost of the three major procedures; see Table \ref{tab6}.

As listed in Table \ref{tab6}, we find that: (1) the first procedure, i.e., the \textit{k}-core decomposition, costs approximately 5\% $\mathrm{\sim}$ 20\% workload; (2) the second procedure, i.e., the motif-based spectral clustering, needs approximately 60\% $\mathrm{\sim}$ 70\% workload; (3) the third procedure, i.e., the label propagation for the remaining nodes, costs 10\% $\mathrm{\sim}$ 20\% workload. In short, the most computationally expensive procedure is the motif-based spectral clustering. Potential work could be conducted in the future to improve the efficiency of this procedure.

\begin{table}[H]
	\small
	\caption{Workload (running time /s) of procedures in the proposed graph clustering algorithm}
	\begin{center}
		\begin{tabular}{|p{0.97in}|p{0.6in}|p{0.9in}|p{0.85in}|p{0.75in}|p{0.56in}|} \hline 
			\textbf{Dataset} & \textbf{\textit{k}-core \newline subgraph} & \textbf{\textit{k}-core \newline decomposition} & \textbf{Motif-based \newline clustering} & \textbf{Recovering} & \textbf{Total} \\ \hline 
			Amazon0302 & 7 & 0.58 & 8.62 & 0.94 & 10.14 \\ \hline 
			web-Stanford & 9 & 0.58 & 16.48 & 1.22 & 18.28 \\ \hline 
			Amazon0505 & 12 & 1.06 & 28.25 & 1.67 & 30.98 \\ \hline 
			web-BerkStan & 7 & 1.16 & 111.36 & 1.76 & 114.28 \\ \hline 
			webGoogle & 28 & 3.65 & 26.94 & 4.18 & 34.77 \\ \hline 
			wikitalk & 10 & 1.37 & 17.61 & 2.51 & 21.49 \\ \hline 
			soc-pokectxt & 40 & 26.45 & 78.00 & 23.19 & 127.64 \\ \hline 
			wiki-topcats & 40 & 20.76 & 58.01 & 23.24 & 102.01 \\ \hline 
			WIKItalk & 10 & 5.37 & 39.19 & 8.38 & 52.94 \\ \hline 
			sx-stackoverflow & 80 & 36.00 & 142.24 & 29.76 & 208.00 \\ \hline 
		\end{tabular}
		\label{tab6}
	\end{center}
\end{table}

\section{Discussion}
\label{sec5}

In this section, we will discuss several critical issues in the proposed graph clustering algorithm. We first evaluate the impact of the values of \textit{k} in the \textit{k}-core decomposition on the accuracy and efficiency of the proposed graph clustering algorithm, and then we suggest the selection of the relatively optimal value of \textit{k}. we also discuss the advantages and shortcomings of the proposed graph clustering algorithm. Finally, we point out our potential future work for further improvement of the proposed graph clustering algorithm.

\subsection{Impact and Selection of the Values of \textit{k} in the \textit{k}-core Decomposition}

In the proposed graph clustering algorithm, the first essential idea is to perform clustering on local \textit{k}-core subgraphs, rather than on the entire graph. We find that the selection of the values of \textit{k} in the \textit{k}-core decomposition can strongly affect the clustering accuracy and efficiency. 

If selecting a large value of \textit{k}, then too many nodes and edges in the original entire graph will be filtered, and too few nodes and edges remain on the \textit{k}-core subgraphs. In this case, the efficiency of the clustering on the \textit{k}-core subgraph may be high, but meanwhile, the accuracy of the clustering on the \textit{k}-core subgraph may be poor. If selecting a small \textit{k}, the above situations would be verse vice. Therefore, a critical question arises in the \textit{k}-core decomposition, i.e., how to determine a relatively optimal value of \textit{k}?

In this subsection, we will try to answer the above question by (1) evaluating the impact of the values of \textit{k} on the efficiency and accuracy of clustering and (2) suggesting the selection of the value of \textit{k}.

\subsubsection{Evaluation of the Impact of \textit{k} on the Clustering Accuracy and Efficiency}

The essential idea behind the proposed graph clustering algorithm is to perform the motif-based spectral clustering on the top \textit{k}-core subgraphs, rather than on the entire graph. Thus, the first critical issue is the \textit{k}-core decomposition. To learn more about the properties of the \textit{k}-core decompositions, we have investigated the distribution of nodal coreness; see Figure \ref{Figure8} and Figure \ref{Figure9}. 

\begin{figure}[htbp] 
	\centering
	\subfigure[26w-Amazon0302]{
		\label{Figure10a}       
		\includegraphics[width=0.481\textwidth]{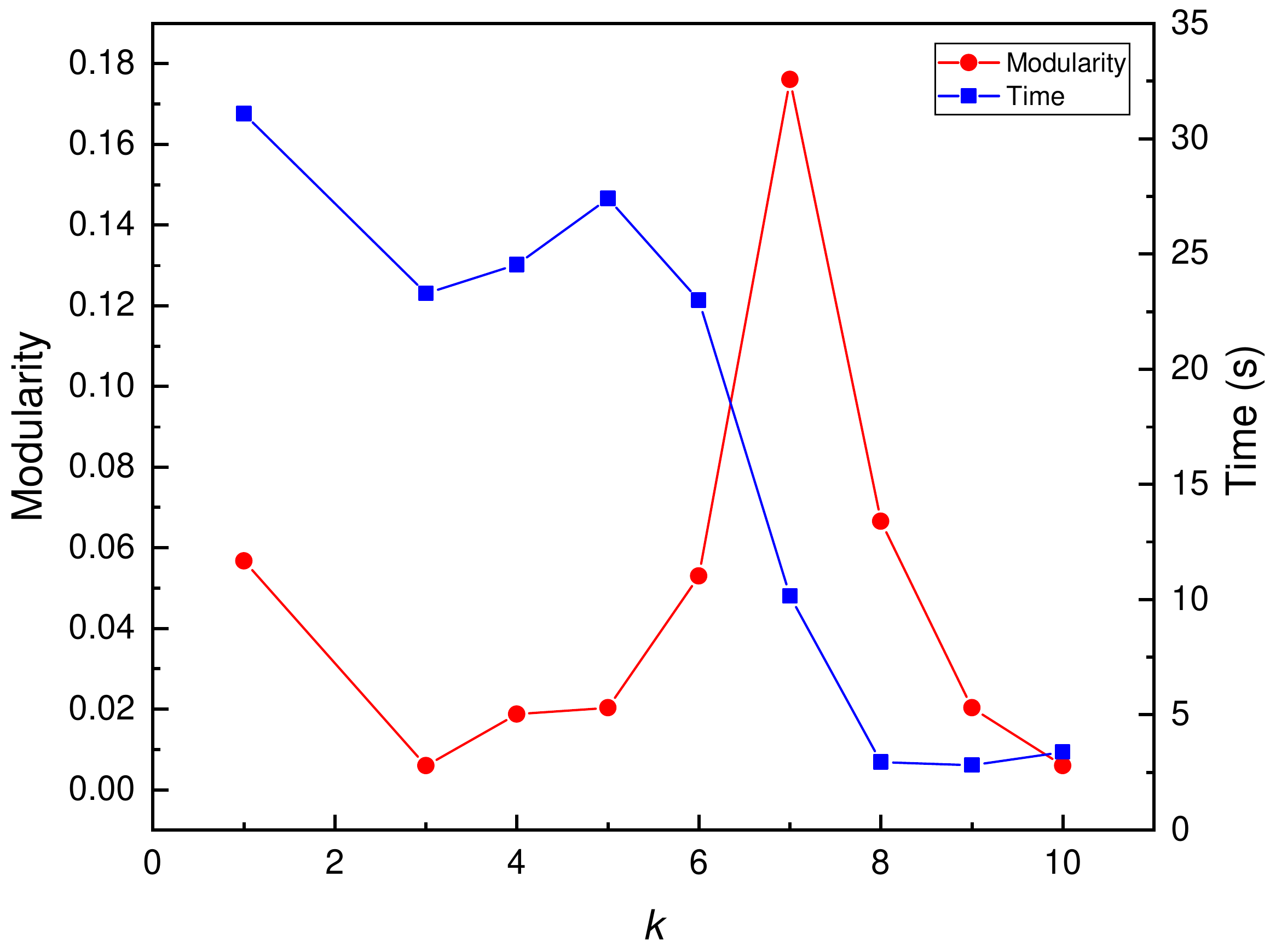}
	}
	\subfigure[28w-web-Stanford]{
		\label{Figure10b}       
		\includegraphics[width=0.481\textwidth]{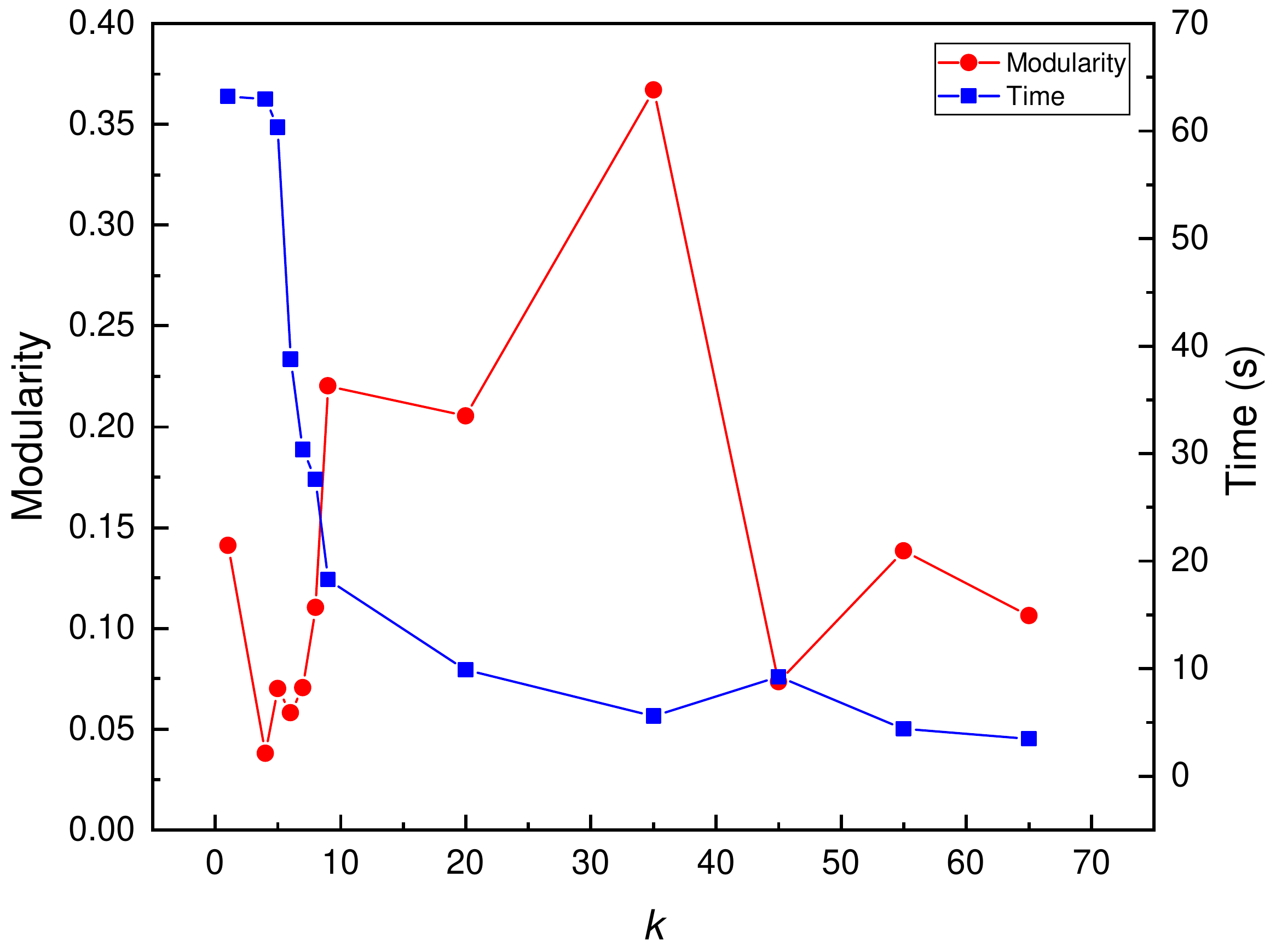}
	}
	\subfigure[41w-Amazon0505]{
		\label{Figure10c}       
		\includegraphics[width=0.481\textwidth]{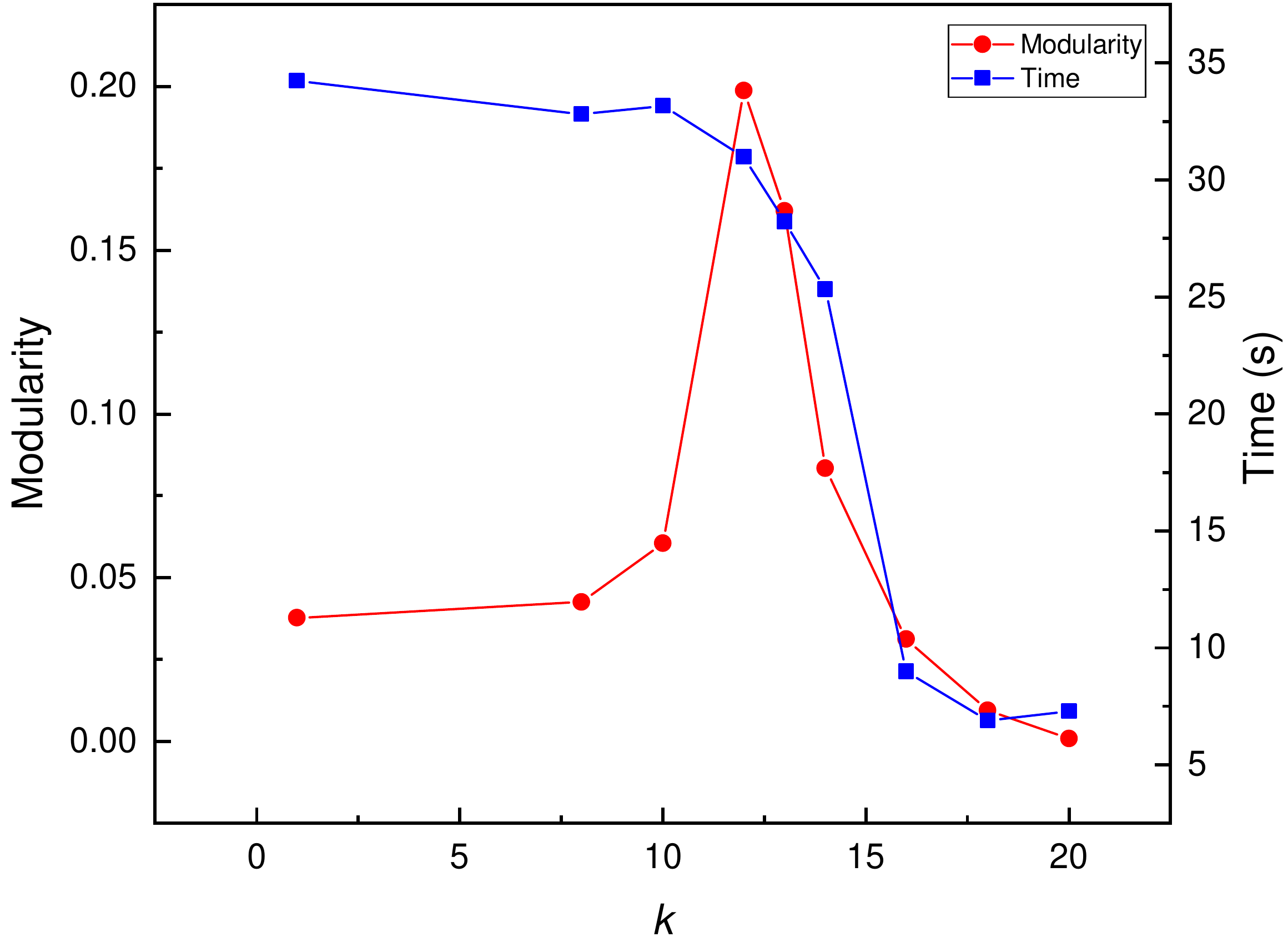}
	}
	\subfigure[68w-web-BerkStan]{
		\label{Figure10d}       
		\includegraphics[width=0.481\textwidth]{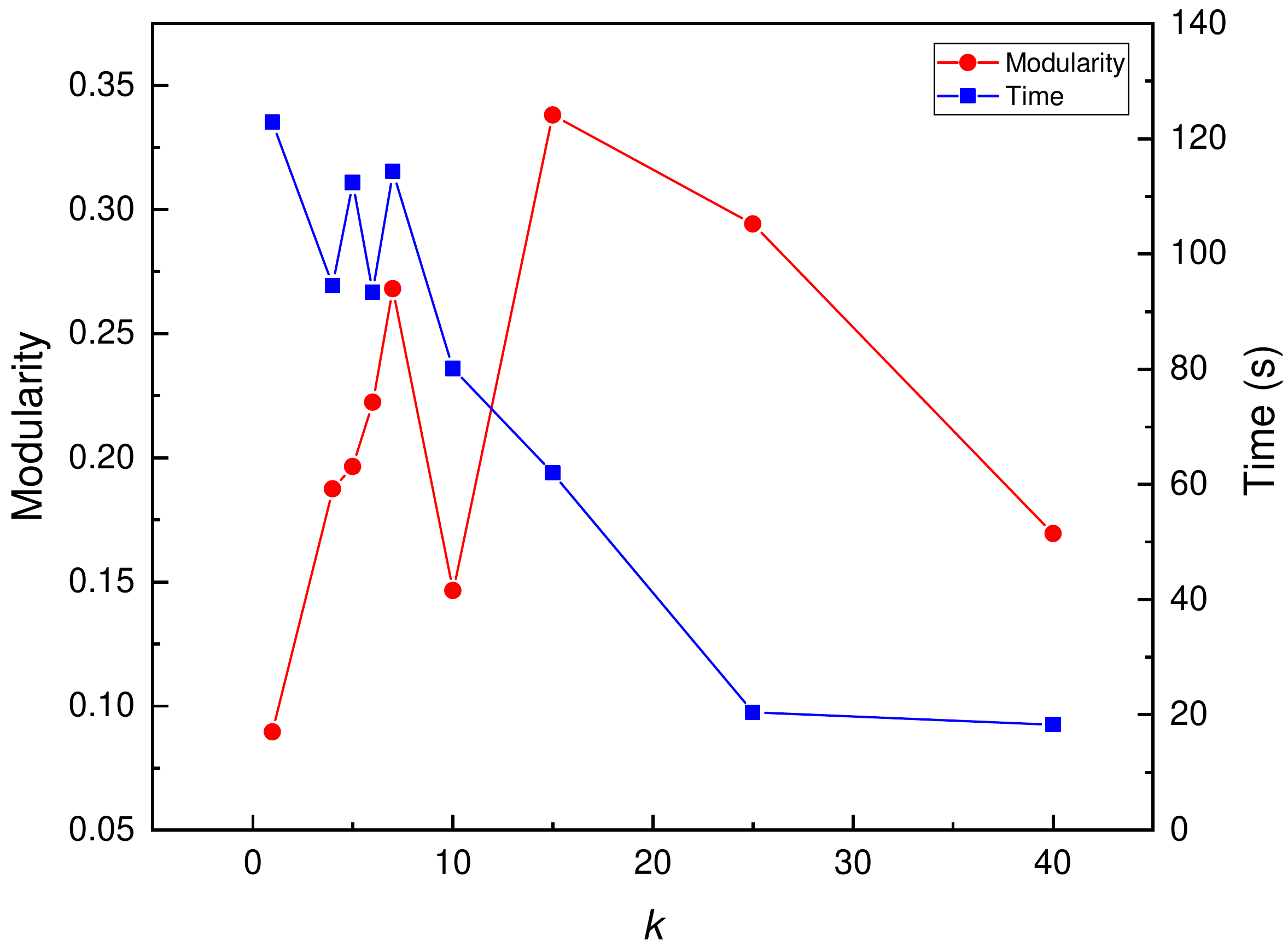}
	}
	\subfigure[170w-wiki-topcats]{
		\label{Figure10e}       
		\includegraphics[width=0.481\textwidth]{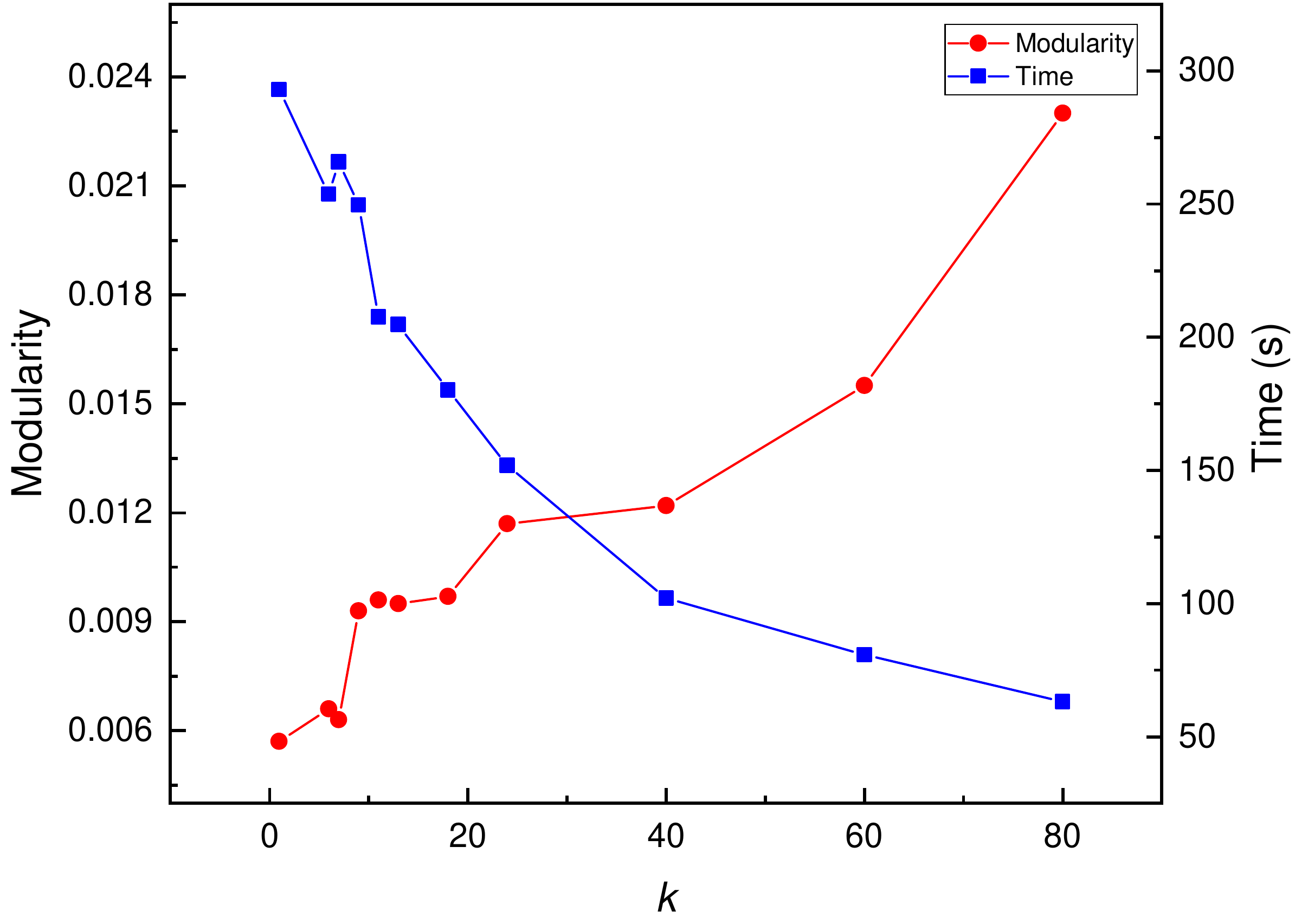}
	}
	\caption{Impact of values of \textit{k} on the efficiency and accuracy for networks with the normal distribution of coreness}
	\label{Figure10}       
\end{figure}

\begin{figure}[htbp] 
	\centering
	\subfigure[87w-webGoogle]{
		\label{Figure11a}       
		\includegraphics[width=0.481\textwidth]{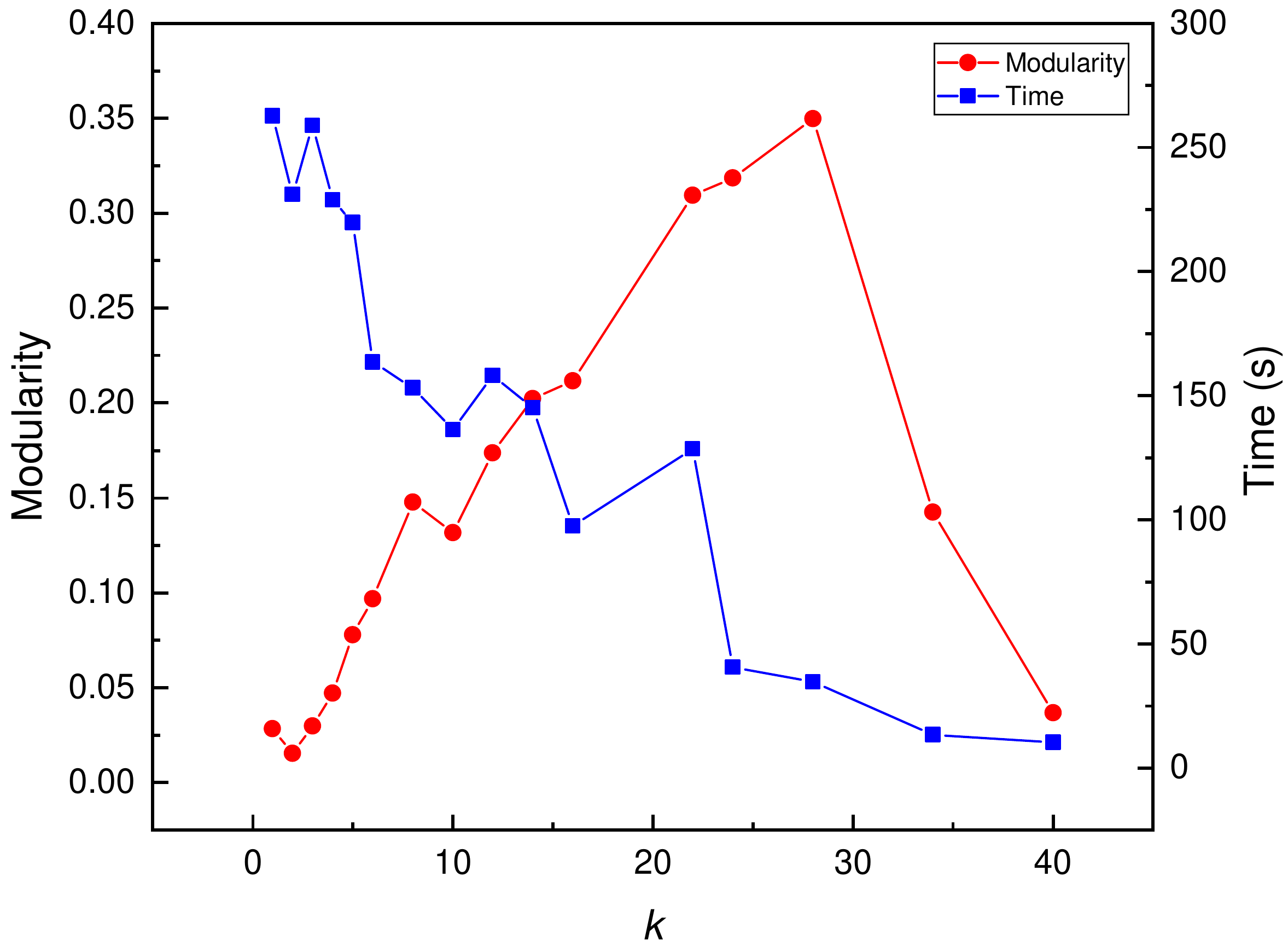}
	}
	\subfigure[110w-wikitalk]{
		\label{Figure11b}       
		\includegraphics[width=0.481\textwidth]{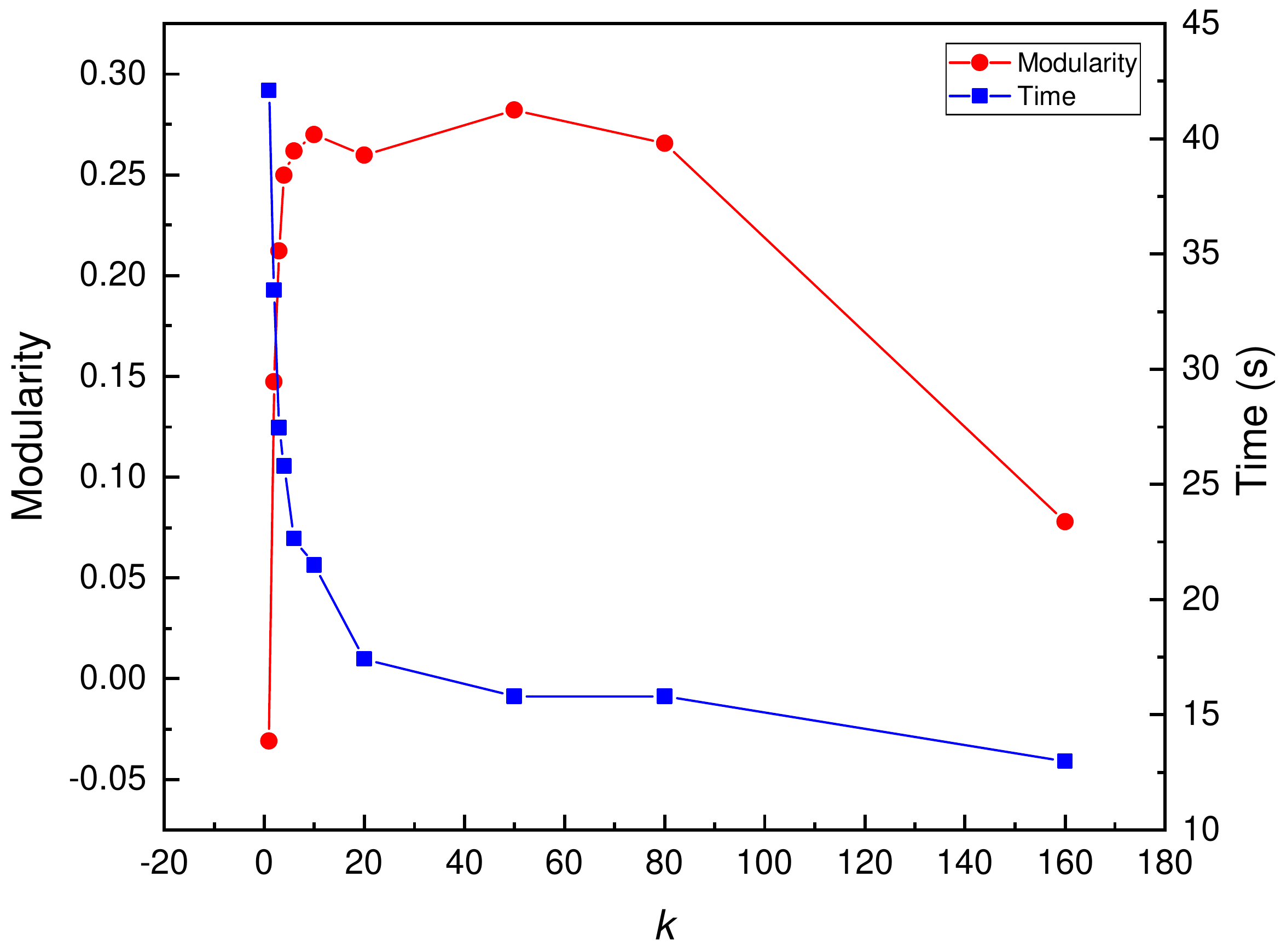}
	}
	\subfigure[160w-soc-pokectxt]{
		\label{Figure11c}       
		\includegraphics[width=0.481\textwidth]{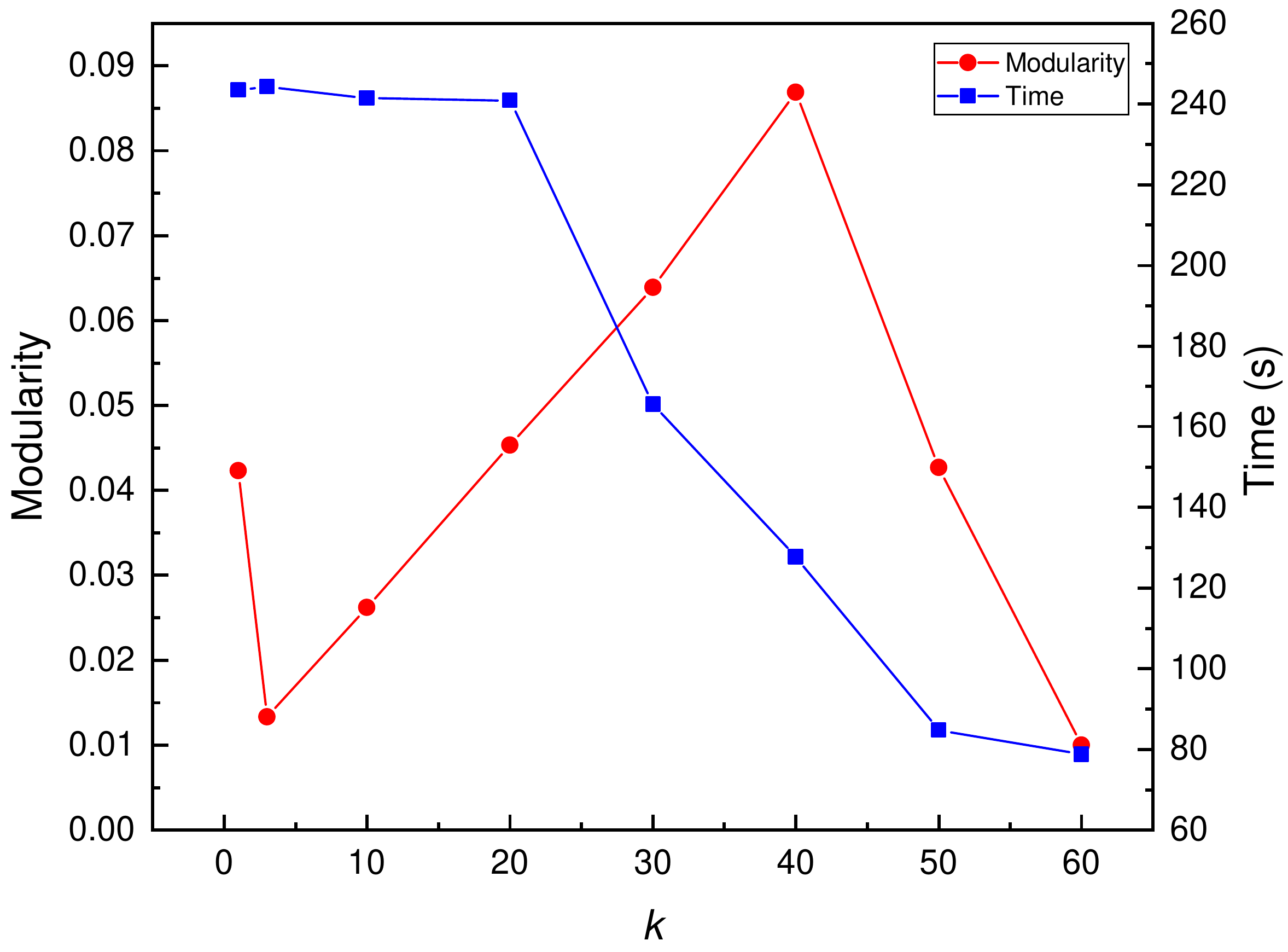}
	}
	\subfigure[230w-WIKItalk]{
		\label{Figure11d}       
		\includegraphics[width=0.481\textwidth]{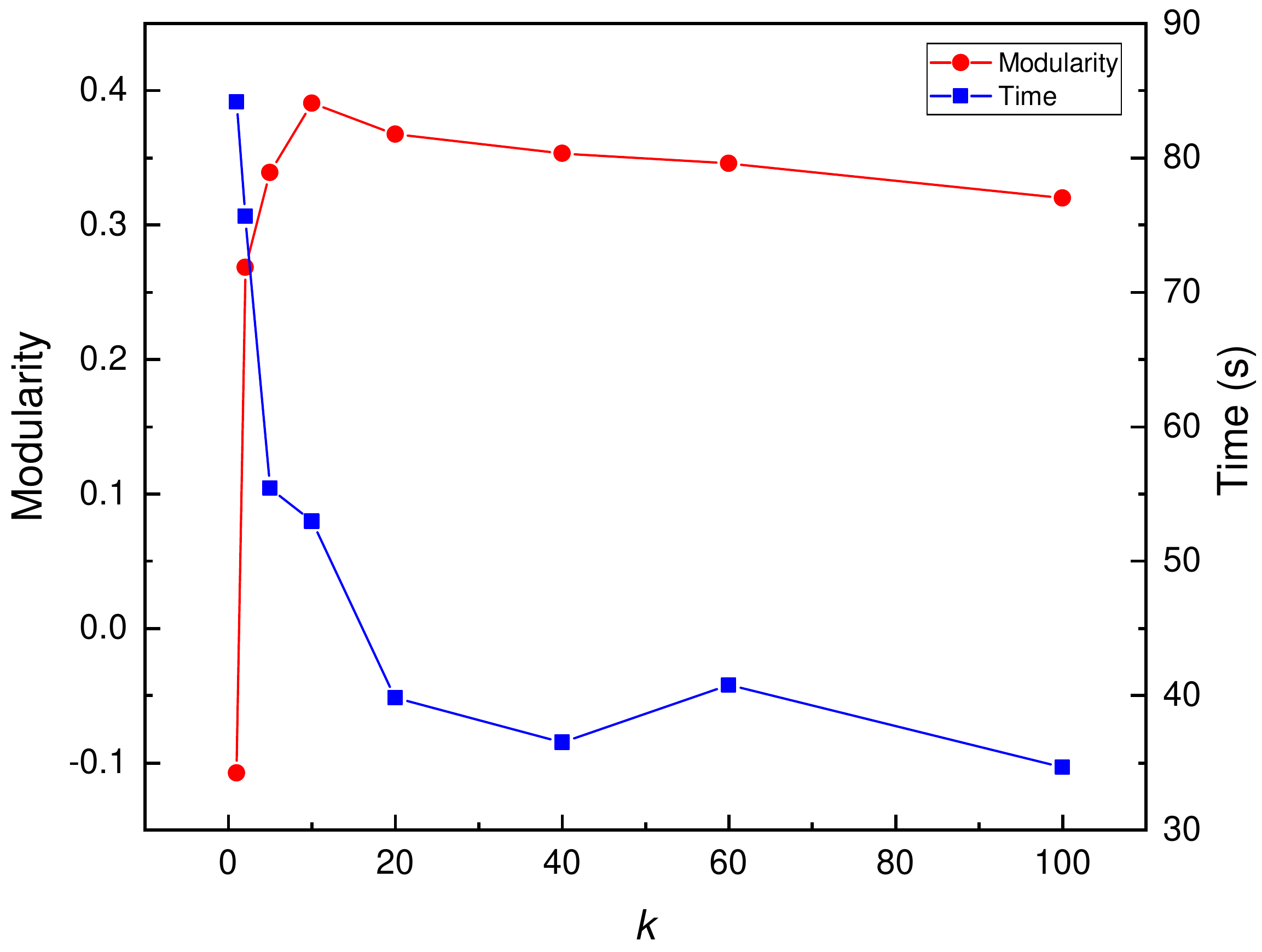}
	}
	\subfigure[250w-sx-stackoverflow]{
		\label{Figure11e}       
		\includegraphics[width=0.481\textwidth]{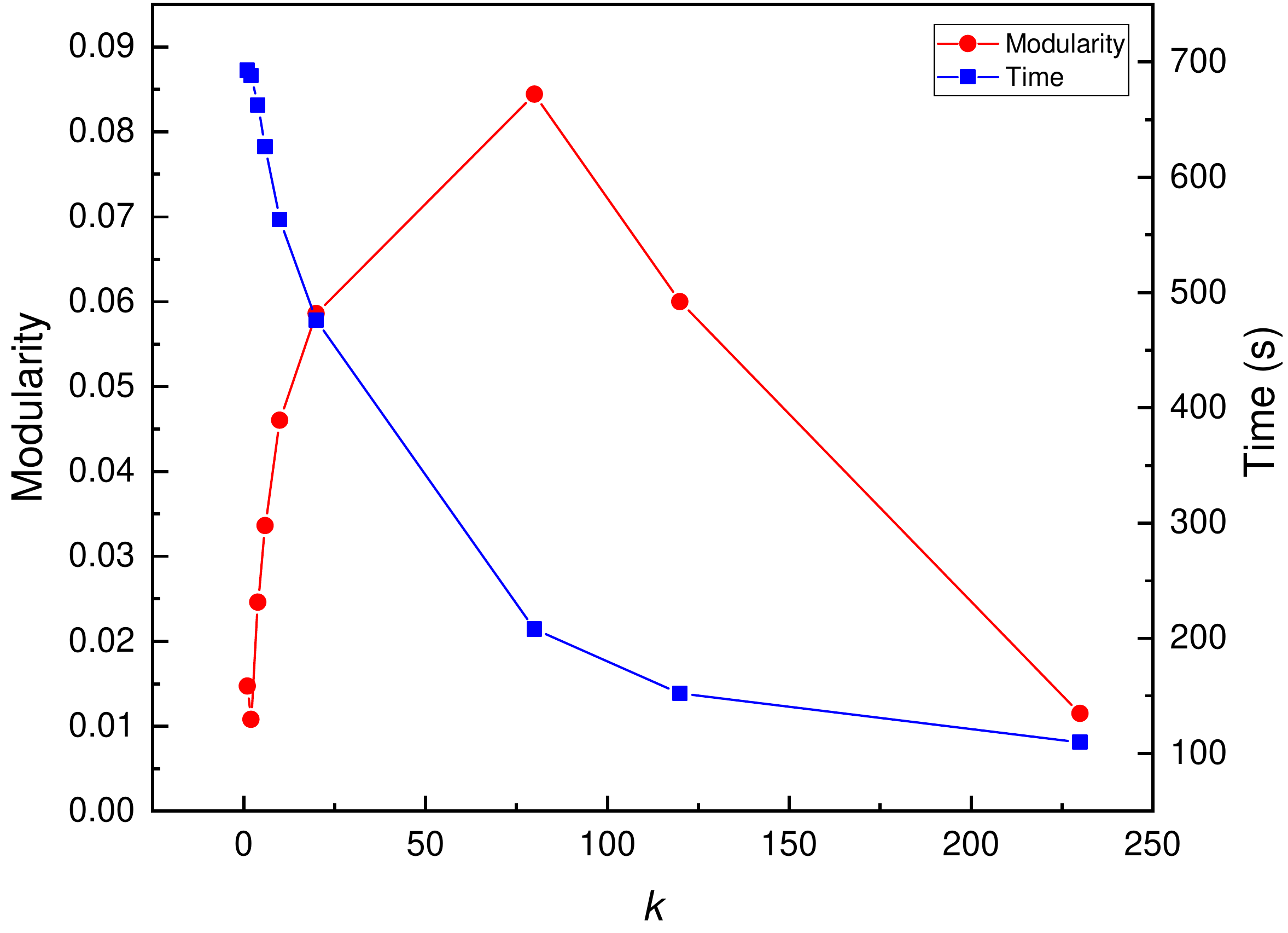}
	}
	\caption{Impact of values of \textit{k} on the efficiency and accuracy for networks with the power-law distribution of coreness}
	\label{Figure11}       
\end{figure}

We have found that, similar to the distribution of nodal degree, the distribution of nodal coreness also follow (1) the normal distribution or (2) the power-law distribution. The above two types of networks are topologically different. And the impact of the values of \textit{k} on the accuracy and efficiency for the above two types of networks may also differ. Thus, we specifically investigate the impact of values of \textit{k} on the accuracy and efficiency of clustering for the above two types of networks; see Figure \ref{Figure10} and Figure \ref{Figure11}.

For the networks with the normal distribution of coreness, we find that: with the increases of \textit{k}, the efficiency of clustering always increase, while the accuracy of clustering first increases, and then decreases when the \textit{k} reaches a specific threshold. For the networks with the power-law distribution of coreness, we can also achieve the above observations. 

However, for the above two types of networks, there is a difference between the trends of the impact of \textit{k} on accuracy. For the networks with the normal distribution of coreness, the accuracy starts to decrease when reaching a relatively “small” value of \textit{k} (Figure \ref{Figure10}). in contrast, for the networks with a power-law distribution of coreness, when the value of \textit{k} becomes very large, the accuracy will decrease (Figure \ref{Figure11}).

To learn more about the aforementioned difference, we investigate the number of nodes on the specified \textit{k}-core subgraph. Obviously, with the increase of the values of \textit{k}, fewer and fewer nodes would be on the \textit{k}-core subgraph. For example, the nodes on the 6-core subgraph are more than that on the 7-core subgraph. We first record the numbers of the entire graph and the number of nodes on the specified and then calculate the percentage of nodes on the \textit{k}-core subgraph over the entire graph. 

We find that: (1) for the networks with the normal distribution of nodal coreness, when the percentage of nodes on the \textit{k}-core subgraph over the entire graph reaches approximately 40\% $\mathrm{\sim}$ 50\%, the accuracy will decreases; (2) for the networks with the power-law distribution of nodal coreness, when the percentage of nodes on the \textit{k}-core subgraph over the entire graph reaches approximately 5\% $\mathrm{\sim}$ 10\%, the accuracy will decreases.

Here we try to explain the reasons why in several cases, the clustering accuracy first increases and then decreases with the increase of the values of \textit{k}. 

Motifs are the blocks consisting of several nodes and directed edges. For example, there are 13 types of three-nodes motifs. A possible phenomenon is that, in the ``dense''  region of a network, the probability of being able to find motifs is higher than that in the ``sparse'' region of the network. In other words, in the ``inner'' region of a network (i.e., the region with higher \textit{k}-core values), more motifs can be found than that in the ``outer'' region of the network (i.e., the region with lower \textit{k}-core values). 

In the motif-based spectral clustering algorithm, the Laplacian matrix is constructed based on network motifs, rather than based on network edges. Thus, the accuracy of the motif-based spectral clustering algorithm is strongly dependent on the number of the probability of being able to find motifs in a given network. If more motifs can be found on the entire network or even in a part of the network, the accuracy of the motif-based spectral clustering would become better. 

Therefore, for a very large network, in the ``outer'' region (i.e.. in the region with lower \textit{k}-core values), not so many motifs can be found. In this case, the probability of being able to find motifs would be low, and the accuracy of the motif-based spectral clustering maybe not satisfying. With the increase of the values of \textit{k}, the probability of being able to find motifs would be higher, and thus in general, the accuracy of the motif-based spectral clustering may increase. However, with a very large value of \textit{k}, the absolute number of motifs that can be found is too small. In this case, the accuracy of the motif-based spectral clustering may decrease.

In summary, generally, with the increase of the values of \textit{k}, the accuracy of the motif-based spectral clustering increase first; then, when the \textit{k} becomes very large, the accuracy decrease. 

Moreover, the motif-based clustering method is quite efficient when clustering the core area of the network with a certain clustering accuracy. This is the reason why the second procedure of the proposed KCoreMotif does not employ common clustering methods such as spectral clustering to cluster the nodes on the top \textit{k}-core subgraph.

\subsubsection{Suggestion of the Value of \textit{k} for the Balanced Accuracy and Efficiency}

As discussed in Subsection 5.1.1, in order to analyze the \textit{k}-core decomposition of networks, we investigated the distribution of the nodal coreness and fitted the curve; see Figure \ref{Figure8} and Figure \ref{Figure9}.

Here we will give our suggestion to the section of the values of \textit{k}. By carefully analyzing the distributions of nodal coreness illustrated in Figure \ref{Figure8} and Figure \ref{Figure9} and the impact of values of \textit{k} presented in Figure \ref{Figure10} and Figure \ref{Figure11}, we find that: 

(1) For the networks with the normal distribution of nodal coreness, if selecting a value of \textit{k} such that approximately 50\% $\mathrm{\sim}$ 60\% nodes of the original network are filtered, i.e., 40\% $\mathrm{\sim}$ 50\% nodes are still on the \textit{k}-core subgraph, then an acceptable balance between the efficiency and accuracy of clustering can be achieved. 

(2) For the networks with the power-law distribution of nodal coreness, if selecting a value of \textit{k} such that approximately 90\% $\mathrm{\sim}$ 95\% nodes of the original network are filtered, i.e., 5\% $\mathrm{\sim}$ 10\% nodes are still on the \textit{k}-core subgraph, then an acceptable balance between the efficiency and accuracy of clustering can be achieved. 

Therefore, in the practical applications, the \textit{k}-core decomposition is first conducted; then, the distribution of coreness should be investigated and fitted to determine whether it is normal or power-law; and finally, the values of \textit{k} can be selected. Note that, according to the above suggestions, the selected value of \textit{k} is ``roughly'' optimal, not always the best.

\subsection{Advantages and Shortcomings of the Proposed Graph Clustering Algorithm}

The major advantage of the proposed graph clustering algorithm is its efficiency while satisfying the accuracy. Compared with the conventional spectral clustering algorithm, two essential ideas are exploited to enhance efficiency, i.e., (1) and (2). Compared with the famous motif-based spectral clustering algorithm proposed by Jure Leskovec \cite{r23}, the benefit of the \textit{k}-core decomposition is utilized.

The major shortcoming of the proposed graph clustering algorithm is the fact that it has three main procedures. That is, it is algorithmically more complex than the conventional spectral clustering algorithm and the famous motif-based spectral clustering algorithm.

Another critical issue is that, in the proposed graph clustering algorithm, the M6 motif is commonly used. And we did not carefully compare and evaluate the impact of the types of motifs on the accuracy and efficiency of the motif-based spectral clustering algorithm. The reason why not conducting such work is that, in the work of the motif-based spectral clustering algorithm proposed by Jure Leskovec \cite{r23}, they have carefully analyzed and evaluate the impact of the types of motifs on the accuracy and efficiency. In the proposed graph clustering algorithm, more specifically, in the second major procedures of the proposed algorithm, the famous motif-based spectral clustering algorithm proposed by is well invoked on \textit{k}-core subgraphs; and we did not revise or modify the ideas or process of the aforementioned motif-based spectral clustering algorithm. 

\subsection{Outlook and Future Work}

As analyzed several times, spectral clustering is inherently difficult to deal with large-scale networks. This is because that in the spectral clustering, it needs to conduct computationally expensive matrix manipulations, e.g., (1) the construction of a large adjacency matrix and Laplacian matrix and (2) the calculation of eigenvectors of the large Laplacian matrix. Even if the large adjacency matrix and Laplacian matrix are stored in compressed and sparse formats such as the COOrdinate (COO) or Compressed Sparse Column (CSC) formats, the manipulations of matrices are quite computationally expensive. 

There are two common strategies to enhance the spectral clustering algorithm. The first strategy for conducting spectral clustering on large networks is to design parallel spectral algorithms on various platforms such as multi-core CPUs, many-core GPUs, and even clouds. The second strategy is to reduce the computational cost of the matrix assembly and manipulation, e.g., the use of the Nyström extension method \cite{r15} for approximating similarity matrix and feature decomposition. 

Therefore, in the future, we will adopt the above two strategies to further improve the computational efficiency of the proposed graph clustering algorithm to work with even larger networks.

\section{Conclusions}
\label{sec6}

In this paper, we have proposed an efficient graph clustering algorithm for large networks by exploiting \textit{k}-core decomposition and motifs. The essential idea behind the proposed clustering algorithm is to perform the motif-based spectral clustering algorithm on \textit{k}-core subgraphs. We first conduct the \textit{k}-core decomposition of the large input network. We then perform the motif-based spectral clustering for the top \textit{k}-core subgraphs; Finally, we group the remaining vertices in the rest (\textit{k}-1)-core subgraphs into previously found clusters and required to obtain the clusters of the large input network. To evaluate the performance of the proposed graph clustering algorithms, we use both the conventional and motif-based spectral clustering algorithms as the baselines and compare our algorithm with them for 18 groups of real-world datasets. Comparative results demonstrate that the proposed clustering algorithm is accurate yet efficient for large networks, which also means that it can be further used to evaluate both the intra-cluster and inter-cluster trusts on large networks.

\section*{Declaration of Competing Interest}
The authors declare that they have no known competing financial interests or personal relationships that could have appeared to influence the work reported in this paper.

\section*{Acknowledgments}
First of all, the authors would like to thank Austin Benson, David Gleich, Jure Leskovec, and Chengbin Peng for providing the source code of their work. This research was jointly supported by the National Natural Science Foundation of China (11602235) and Fundamental Research Funds for China Central Universities (2652018091). The authors would like to thank the editor and the reviewers for their valuable comments.

\section*{Reference}
  \bibliographystyle{elsarticle-num} 
  \bibliography{myRef}

\begin{thebibliography}{10}
\expandafter\ifx\csname url\endcsname\relax
  \def\url#1{\texttt{#1}}\fi
\expandafter\ifx\csname urlprefix\endcsname\relax\def\urlprefix{URL }\fi
\expandafter\ifx\csname href\endcsname\relax
  \def\href#1#2{#2} \def\path#1{#1}\fi

\bibitem{r01}
Z.-J. Du, H.-Y. Luo, X.-D. Lin, S.-M. Yu, A trust-similarity analysis-based
  clustering method for large-scale group decision-making under a social
  network, Information Fusion 63 (2020) 13--29.
\newblock \href {http://dx.doi.org/10.1016/j.inffus.2020.05.004}
  {\path{doi:10.1016/j.inffus.2020.05.004}}.

\bibitem{r02}
W.~Sherchan, S.~Nepal, C.~Paris, A survey of trust in social networks, ACM
  Computing Surveys 45~(4).
\newblock \href {http://dx.doi.org/10.1145/2501654.2501661}
  {\path{doi:10.1145/2501654.2501661}}.

\bibitem{r03}
W.~Jiang, G.~Wang, M.~Bhuiyan, J.~Wu, Understanding graph-based trust
  evaluation in online social networks: Methodologies and challenges, ACM
  Computing Surveys 49~(1).
\newblock \href {http://dx.doi.org/10.1145/2906151}
  {\path{doi:10.1145/2906151}}.

\bibitem{r04}
J.-H. Cho, A.~Swami, I.-R. Chen, A survey on trust management for mobile ad hoc
  networks, IEEE Communications Surveys and Tutorials 13~(4) (2011) 562--583.
\newblock \href {http://dx.doi.org/10.1109/SURV.2011.092110.00088}
  {\path{doi:10.1109/SURV.2011.092110.00088}}.

\bibitem{r05}
K.~Govindan, P.~Mohapatra, Trust computations and trust dynamics in mobile
  adhoc networks: A survey, IEEE Communications Surveys and Tutorials 14~(2)
  (2012) 279--298.
\newblock \href {http://dx.doi.org/10.1109/SURV.2011.042711.00083}
  {\path{doi:10.1109/SURV.2011.042711.00083}}.

\bibitem{r06}
S.~Fortunato, Community detection in graphs, Physics Reports 486~(3-5) (2010)
  75--174.
\newblock \href {http://dx.doi.org/10.1016/j.physrep.2009.11.002}
  {\path{doi:10.1016/j.physrep.2009.11.002}}.

\bibitem{r07}
S.~Fortunato, D.~Hric, Community detection in networks: A user guide, Physics
  Reports 659 (2016) 1--44.
\newblock \href {http://dx.doi.org/10.1016/j.physrep.2016.09.002}
  {\path{doi:10.1016/j.physrep.2016.09.002}}.

\bibitem{r08}
F.~Malliaros, M.~Vazirgiannis, Clustering and community detection in directed
  networks: A survey, Physics Reports 533~(4) (2013) 95--142.
\newblock \href {http://dx.doi.org/10.1016/j.physrep.2013.08.002}
  {\path{doi:10.1016/j.physrep.2013.08.002}}.

\bibitem{r09}
U.~Von~Luxburg, A tutorial on spectral clustering, Statistics and Computing
  17~(4) (2007) 395--416.
\newblock \href {http://dx.doi.org/10.1007/s11222-007-9033-z}
  {\path{doi:10.1007/s11222-007-9033-z}}.

\bibitem{r10}
Z.~Huo, G.~Mei, G.~Casolla, F.~Giampaolo, Designing an efficient parallel
  spectral clustering algorithm on multi-core processors in julia, Journal of
  Parallel and Distributed Computing 138 (2020) 211--221.
\newblock \href {http://dx.doi.org/10.1016/j.jpdc.2020.01.003}
  {\path{doi:10.1016/j.jpdc.2020.01.003}}.

\bibitem{r11}
K.~Gao, G.~Mei, F.~Piccialli, S.~Cuomo, J.~Tu, Z.~Huo, Julia language in
  machine learning: Algorithms, applications, and open issues, Computer Science
  Review 37 (2020) 100254.
\newblock \href
  {http://dx.doi.org/https://doi.org/10.1016/j.cosrev.2020.100254}
  {\path{doi:https://doi.org/10.1016/j.cosrev.2020.100254}}.

\bibitem{r12}
Y.~Jin, J.~Jaja, A high performance implementation of spectral clustering on
  cpu-gpu platforms, 2016, pp. 825--834.
\newblock \href {http://dx.doi.org/10.1109/IPDPSW.2016.79}
  {\path{doi:10.1109/IPDPSW.2016.79}}.

\bibitem{r13}
W.-Y. Chen, Y.~Song, H.~Bai, C.-J. Lin, E.~Chang, Parallel spectral clustering
  in distributed systems, IEEE Transactions on Pattern Analysis and Machine
  Intelligence 33~(3) (2011) 568--586.
\newblock \href {http://dx.doi.org/10.1109/TPAMI.2010.88}
  {\path{doi:10.1109/TPAMI.2010.88}}.

\bibitem{r14}
R.~Jin, C.~Kou, R.~Liu, Y.~Li, Efficient parallel spectral clustering algorithm
  design for large data sets under cloud computing environment, Journal of
  Cloud Computing 2~(1).
\newblock \href {http://dx.doi.org/10.1186/2192-113X-2-18}
  {\path{doi:10.1186/2192-113X-2-18}}.

\bibitem{r15}
C.~Fowlkes, S.~Belongie, F.~Chung, J.~Malik, Spectral grouping using the
  nyström method, IEEE Transactions on Pattern Analysis and Machine
  Intelligence 26~(2) (2004) 214--225.
\newblock \href {http://dx.doi.org/10.1109/TPAMI.2004.1262185}
  {\path{doi:10.1109/TPAMI.2004.1262185}}.

\bibitem{r16}
K.~Zhang, J.~Kwok, Clustered nyström method for large scale manifold learning
  and dimension reduction, IEEE Transactions on Neural Networks 21~(10) (2010)
  1576--1587.
\newblock \href {http://dx.doi.org/10.1109/TNN.2010.2064786}
  {\path{doi:10.1109/TNN.2010.2064786}}.

\bibitem{r17}
M.-A. Belabbas, P.~Wolfe, Spectral methods in machine learning and new
  strategies for very large datasets, Proceedings of the National Academy of
  Sciences of the United States of America 106~(2) (2009) 369--374.
\newblock \href {http://dx.doi.org/10.1073/pnas.0810600105}
  {\path{doi:10.1073/pnas.0810600105}}.

\bibitem{r18}
X.~Zhang, L.~Zong, Q.~You, X.~Yong, Sampling for nyström extension-based
  spectral clustering: Incremental perspective and novel analysis, ACM
  Transactions on Knowledge Discovery from Data 11~(1).
\newblock \href {http://dx.doi.org/10.1145/2934693}
  {\path{doi:10.1145/2934693}}.

\bibitem{r19}
L.~Wang, J.~Bezdek, C.~Leckie, R.~Kotagiri, Selective sampling for approximate
  clustering of very large data sets, International Journal of Intelligent
  Systems 23~(3) (2008) 313--331.
\newblock \href {http://dx.doi.org/10.1002/int.20268}
  {\path{doi:10.1002/int.20268}}.

\bibitem{r20}
S.~Kumar, M.~Mohri, A.~Talwalkar, On sampling-based approximate spectral
  decomposition, 2009, pp. 553--560.

\bibitem{r21}
C.~{Peng}, T.~G. {Kolda}, A.~{Pinar}, {Accelerating Community Detection by
  Using K-core Subgraphs}, arXiv e-prints (2014) arXiv:1403.2226\href
  {http://arxiv.org/abs/1403.2226} {\path{arXiv:1403.2226}}.

\bibitem{r22}
C.~{Giatsidis}, F.~D. {Malliaros}, N.~{Tziortziotis}, C.~{Dhanjal},
  E.~{Kiagias}, D.~M. {Thilikos}, M.~{Vazirgiannis}, {A k-core Decomposition
  Framework for Graph Clustering}, arXiv e-prints (2016) arXiv:1607.02096\href
  {http://arxiv.org/abs/1607.02096} {\path{arXiv:1607.02096}}.

\bibitem{r23}
A.~Benson, D.~Gleich, J.~Leskovec, Higher-order organization of complex
  networks, Science 353~(6295) (2016) 163--166.
\newblock \href {http://dx.doi.org/10.1126/science.aad9029}
  {\path{doi:10.1126/science.aad9029}}.

\bibitem{r24}
U.~Von~Luxburg, A tutorial on spectral clustering, Statistics and Computing
  17~(4) (2007) 395--416.
\newblock \href {http://dx.doi.org/10.1007/s11222-007-9033-z}
  {\path{doi:10.1007/s11222-007-9033-z}}.

\bibitem{r25}
M.~Saerens, F.~Fouss, L.~Yen, P.~Dupont, The principal components analysis of a
  graph, and its relationships to spectral clustering, Vol. 3201, 2004, pp.
  371--383.
\newblock \href {http://dx.doi.org/10.1007/978-3-540-30115-8_35}
  {\path{doi:10.1007/978-3-540-30115-8_35}}.

\bibitem{r26}
J.~Shi, J.~Malik, Normalized cuts and image segmentation, IEEE Transactions on
  Pattern Analysis and Machine Intelligence 22~(8) (2000) 888--905, cited By
  9816.
\newblock \href {http://dx.doi.org/10.1109/34.868688}
  {\path{doi:10.1109/34.868688}}.

\bibitem{r27}
Z.~Wu, An optimal graph theoretic approach to data clustering: Theory and its
  application to image segmentation, IEEE Transactions on Pattern Analysis and
  Machine Intelligence 15~(11) (1993) 1101--1113.
\newblock \href {http://dx.doi.org/10.1109/34.244673}
  {\path{doi:10.1109/34.244673}}.

\bibitem{r28}
S.~Sarkar, P.~Soundararajan, Supervised learning of large perceptual
  organization: Graph spectral partitioning and learning automata, IEEE
  Transactions on Pattern Analysis and Machine Intelligence 22~(5) (2000)
  504--525.
\newblock \href {http://dx.doi.org/10.1109/34.857006}
  {\path{doi:10.1109/34.857006}}.

\bibitem{r29}
L.~Hagen, A.~Kahng, New spectral methods for ratio cut partitioning and
  clustering, IEEE Transactions on Computer-Aided Design of Integrated Circuits
  and Systems 11~(9) (1992) 1074--1085.
\newblock \href {http://dx.doi.org/10.1109/43.159993}
  {\path{doi:10.1109/43.159993}}.

\bibitem{r30}
C.~Ding, X.~He, H.~Zha, M.~Gu, H.~Simon, A min-max cult algorithm for graph
  partitioning and data clustering, 2001, pp. 107--114.

\bibitem{r31}
D.~Arthur, S.~Vassilvitskii, K-means++: The advantages of careful seeding, Vol.
  07-09-January-2007, 2007, pp. 1027--1035.

\bibitem{r32}
S.~Seidman, Network structure and minimum degree, Social Networks 5~(3) (1983)
  269--287.
\newblock \href {http://dx.doi.org/10.1016/0378-8733(83)90028-X}
  {\path{doi:10.1016/0378-8733(83)90028-X}}.

\bibitem{r33}
W.~Pan, B.~Li, J.~Liu, Y.~Ma, B.~Hu, Analyzing the structure of java software
  systems by weighted k-core decomposition, Future Generation Computer Systems
  83 (2018) 431--444.
\newblock \href {http://dx.doi.org/10.1016/j.future.2017.09.039}
  {\path{doi:10.1016/j.future.2017.09.039}}.

\bibitem{r34}
M.~Al-garadi, K.~Varathan, S.~Ravana, Identification of influential spreaders
  in online social networks using interaction weighted k-core decomposition
  method, Physica A: Statistical Mechanics and its Applications 468 (2017)
  278--288.
\newblock \href {http://dx.doi.org/10.1016/j.physa.2016.11.002}
  {\path{doi:10.1016/j.physa.2016.11.002}}.

\bibitem{r35}
F.~Morone, G.~Del~Ferraro, H.~Makse, The k-core as a predictor of structural
  collapse in mutualistic ecosystems, Nature Physics 15~(1) (2019) 95--102.
\newblock \href {http://dx.doi.org/10.1038/s41567-018-0304-8}
  {\path{doi:10.1038/s41567-018-0304-8}}.

\bibitem{r36}
X.~He, H.~Zhao, W.~Cai, G.-G. Li, F.-D. Pei, Analyzing the structure of
  earthquake network by k-core decomposition, Physica A: Statistical Mechanics
  and its Applications 421 (2015) 34--43.
\newblock \href {http://dx.doi.org/10.1016/j.physa.2014.11.022}
  {\path{doi:10.1016/j.physa.2014.11.022}}.

\bibitem{r37}
V.~{Batagelj}, M.~{Zaversnik}, {An O(m) Algorithm for Cores Decomposition of
  Networks}, arXiv e-prints (2003) cs/0310049\href
  {http://arxiv.org/abs/cs/0310049} {\path{arXiv:cs/0310049}}.

\bibitem{r38}
O.~Zoidi, E.~Fotiadou, N.~Nikolaidis, I.~Pitas, Graph-based label propagation
  in digital media: A review, ACM Computing Surveys 47~(3).
\newblock \href {http://dx.doi.org/10.1145/2700381}
  {\path{doi:10.1145/2700381}}.

\bibitem{r39}
S.~Garza, S.~Schaeffer, Community detection with the label propagation
  algorithm: A survey, Physica A: Statistical Mechanics and its Applications
  534.
\newblock \href {http://dx.doi.org/10.1016/j.physa.2019.122058}
  {\path{doi:10.1016/j.physa.2019.122058}}.

\bibitem{r40}
M.~Newman, Modularity and community structure in networks, Proceedings of the
  National Academy of Sciences of the United States of America 103~(23) (2006)
  8577--8582.
\newblock \href {http://dx.doi.org/10.1073/pnas.0601602103}
  {\path{doi:10.1073/pnas.0601602103}}.

\end{thebibliography}
  

\end{document}